\documentclass{pasj00}

\begin{document}
\SetRunningHead{Y. Takeda et al.}{Li, C, and O Abundances of Hyades F--G Stars}
\Received{2012/09/10}
\Accepted{2012/12/20}
\title{Lithium, Carbon, and Oxygen Abundances of \\
Hyades F--G Type Stars
\thanks{Based on data collected at Okayama Astrophysical Observatory (NAOJ, Japan).
}}
\author{Yoichi \textsc{Takeda}}
\affil{National Astronomical Observatory, 2-21-1 Osawa, Mitaka, Tokyo 181-8588}
\email{takeda.yoichi@nao.ac.jp}

\author{Satoshi \textsc{Honda}}
\affil{Nishi-Harima Astronomical Observatory, Center for Astronomy, \\
University of Hyogo, 407-2, Nishigaichi, Sayo-cho, Sayo, Hyogo 679-5313}
\email{honda@nhao.jp}

\author{Takashi \textsc{Ohnishi}}
\affil{Nagoya City Science Museum, 17-1, Sakae 2-chome, Naka-ku, Nagoya 460-0008}
\email{ohnishi@nagoya-p.jp}

\author{Michiko \textsc{Ohkubo}, Ryuko \textsc{Hirata}}
\affil{Department of Astronomy, Faculty of Science, Kyoto University, Sakyo-ku, Kyoto 606-8502}
\email{michiko@kusastro.kyoto-u.ac.jp, hirata@kusastro.kyoto-u.ac.jp}

\and

\author{Kozo \textsc{Sadakane}}
\affil{Astronomical Institute, Osaka Kyoiku University, Asahigaoka, Kashiwara, Osaka 582-8582}
\email{sadakane@cc.osaka-kyoiku.ac.jp}

\maketitle

\begin{abstract}

In an attempt to carry out a systematic study on the behavior of 
the photospheric abundances of Li, C, and O (along with Fe) 
for Hyades main-sequence stars in the $T_{\rm eff}$ range 
of $\sim$~5000--7000~K, we conducted an extensive 
spectrum-synthesis analysis applied to four spectral regions  
(comprising lines of Fe-group elements, Li~{\sc i} 6708 line,
C~{\sc i} 7111--7119 lines, and O~{\sc i} 6156--8 lines)
based on the high-dispersion spectra of 68 selected F--G type 
stars belonging to this cluster.
The abundances of C and O turned out to be fairly uniform
in a marginally supersolar level such like the case of Fe:
$\langle$[C/H]$\rangle = +0.15$~$(\sigma = 0.08)$,
$\langle$[O/H]$\rangle = +0.22$~$(\sigma = 0.14)$, and
$\langle$[Fe/H]$\rangle = +0.11$~$(\sigma = 0.08)$,
suggesting that the primordial abundances are almost retained 
for these elements. Strictly, however, they show a slightly
increasing trend with a decrease in $T_{\rm eff}$
(typically on the order of $\sim 10^{-4}$~dex~K$^{-1}$); while 
this might be due to an improper choice of atmospheric parameters, 
we found it hard to give a quantitatively reasonable explanation. 
Regarding Li, we confirmed the well-known $T_{\rm eff}$-dependent 
trend in the Li abundance reported so far (a conspicuous Li-trough at 
6300~K~$\ltsim T_{\rm eff} \ltsim 6700$~K and a progressive decrease
toward a lower $T_{\rm eff}$ at $T_{\rm eff} \ltsim 6000$~K), which means
that the surface Li of Hyades stars is essentially controlled only by 
$T_{\rm eff}$ and other parameters such as the rotational velocity
are almost irrelevant.
\end{abstract}


\section{Introduction}

\subsection{Chemical Abundances of Hyades Cluster Stars}

Since stars belonging to a cluster are considered to have formed 
almost at the same time out of chemically near-homogeneous material,
spectroscopically studying the photospheric abundances of
cluster stars can provide us with valuable information on 
the primordial gas (chemical composition, degree of homogeneity, etc.) 
as well as physical processes that may posteriori affect surface 
abundances (e.g., mixing/segregation in the stellar envelope). 

Above all, the Hyades cluster (comprising A--M stars, its age and 
distance being precisely determined as $6.25 \times 10^{8}$~yr and 47~pc,
respectively; cf. Perrymann et al. 1998; its metallicity is known 
to be slightly supersolar at [Fe/H]\footnote{
[X/H] means the differential abundance of element X of a star 
relative to that of the Sun; i.e.,
[X/H]~$\equiv A({\rm X})_{*} - A({\rm X})_{\odot}$.
Here, $A$(X) denotes the logarithmic (number) abundance of an element X
with the usual normalization of $A$(H) = 12.00.}
$\sim$~0.1--0.2; cf. Takeda 2008) 
is one of the most suitable galactic open clusters for this objective 
because of its proximity and well established parameters of member stars.

The purpose of this article is to report the results of our new 
systematic abundance studies on Li (a key element whose abundance 
reflects mixing history of the envelope because of its fragile 
nature) as well as C and O (most abundant metals playing important 
roles in the galactic chemical evolution) for a number of F--G type 
stars of the Hyades cluster.

\subsection{Oxygen}

Oxygen abundances of Hyades main-sequence stars are not yet 
sufficiently well understood. To describe the situation on 
this matter, stars have to be divided into two groups to be 
separately treated; A-type stars ($T_{\rm eff} \gtsim 7000$~K) 
and FGK-type stars ($T_{\rm eff} \ltsim$~7000~K).

\subsubsection{[O/H] in Hyades A stars}

While it is certain that Hyades A-type stars (including Am stars)
in the $T_{\rm eff}$ range of $\sim$~7000--9000~K show diversified 
O-abundances in their photospheres (being anti-correlated 
with Fe in the sense that a deficit of O is accompanied by 
an excess of Fe), presumably due to the process of 
chemical segregation (atomic diffusion) in the stellar envelope
(e.g., Richer et al. 2000), the issue of how this anomaly 
(Am peculiarity) is triggered is still somewhat controversial 
(i.e., only the rotational velocity is responsible? or some 
other factors are involved?).

Takeda and Sadakane (1997) reported the existence of a  
$v_{\rm e}\sin i$-dependence of [O/H] (a positive correlation at 
$v_{\rm e}\sin i \ltsim 100$~km~s$^{-1}$) from their spectrum-fitting 
analysis of the O~{\sc i}~7771--5 triplet lines on 18 Hyades A-type
stars.

However, Varenne and Monier (1999) did not corroborate 
this relation in their analysis for 19 Hyades A dwarfs
using O~{\sc i}~6155--8 lines, though a tendency of lower [O/H] 
for A-type slower rotators ($v_{\rm e}\sin i \ltsim 100$~km~s$^{-1}$) 
is observed (cf. their figure 5 therein).

Nevertheless, Takeda et al. (2009) reconfirmed the clear 
rotation-dependent trend of [O/H] in 23 Hyades A-type stars
(i.e., an increasing tendency from $\sim -0.5$ to 
$\sim 0.0$ with an increase in $v_{\rm e}\sin i$ at 
0~km~s$^{-1} \ltsim v_{\rm e}\sin i \ltsim 100$~km~s$^{-1}$,
while an almost constant [O/H] of $\simeq 0.05\pm 0.10$ at 
$v_{\rm e}\sin i \gtsim 100$~km~s$^{-1}$; cf. figure 8a therein) 
based on the O~{\sc i}~6155--8 lines as used by Varenne and 
Monier (1999). 

Yet, Gebran et al. (2010) concluded in their reanalysis of
16 Hyades A-stars that any meaningful correlation does not exist 
between [O/H] and $v_{\rm e}\sin i$, again the same conclusion 
as that of Varenne and Monier (1999). 
However, since their figure 7 apparently exhibits a trend of 
subsolar [O/H] for A-stars of lower $v_{\rm e}\sin i$, this might
rather be a matter of {\it definition} in their using the word of
`correlation' or `dependence.'

\subsubsection{[O/H] in Hyades FGK stars}

Meanwhile, when it comes to discussing the primordial O-abundance
of this cluster, it is necessary to establish the precise [O/H] 
values of unevolved F--G--K dwarfs. Unfortunately, however, despite 
a number of investigations done so far, any consensus has not yet 
been accomplished: 
\begin{itemize}
\item Tomkin and Lambert (1978) studied two Hyades F-type stars
(45 Tau = HD~26462 and HD~27561) using O~{\sc i} 9260--9266 lines 
and obtained [O/H] = +0.18 and +0.02, respectively.
\item Garc\'{\i}a-Lop\'{e}z et al. (1993) reported 
from their analysis of O~{\sc i} 7771-5 lines that the 
mean [O/H] of F-type stars is slightly subsolar ($-0.05$ or $-0.10$ 
depending on the sample selection).
\item King (1993) concluded based on O~{\sc i}~7771--5 lines
that mean [O/H] of four Hyades late-F stars is supersolar ([O/H] = 0.27).
\item King and Hiltgen's (1996) analysis of [O~{\sc i}]~6300 line on 
two Hyades early K dwarfs yielded [O/H]~$\sim +0.15$.
\item Takeda et al. (1998) derived the mean [O/H] of $\sim +0.1$ 
for 11 Hyades F stars based on the O~{\sc i}~8446 line.
\item Analysis of F-type stars (7000~K~$\gtsim T_{\rm eff} \gtsim 6000$~K) 
by Varenne and Monier (1999) suggested a marginally supersolar trend
of [O/H] ranging from 0.0 to +0.3.
\item Schuler et al. (2006a,b) made an extensive O-abundance study for
many Hyades G--K dwarfs by using O~{\sc i} 7771--5, [O~{\sc i}]~6300,
and CO lines; they found a considerable increase of supersolar 
[O/H] (especially for those from O~{\sc i} 7771--5 lines)  
toward decreasing $T_{\rm eff}$ at $T_{\rm eff} \ltsim 5500$~K
(presumably due to the enhanced chromospheric activity which makes 
the classical treatment inapplicable), and could not accomplish 
any consistent solution for the oxygen abundance of the cluster.
\item Gebran et al.'s (2010) analysis of F-type stars 
(7000~K~$\gtsim T_{\rm eff} \gtsim 6000$~K) implied a rather large 
diversity (around [O/H]~$\sim 0$) amounting to $\sim \pm 0.4$~dex.
\end{itemize}

Given this complicated situation, a new comprehensive study may be 
worth carrying out. So far, we have been involved with investigating 
the oxygen abundances of field main-sequence stars by using the 
spectrum-fitting technique applied to O~{\sc i}~6156--8 lines 
(Takeda et al. 1999 [late B and A stars]; 
Takeda \& Honda 2005 [from late F to early K stars]; Takeda et al. 
2010 [B stars]).
These high-excitation permitted O~{\sc i} lines are regarded to be
well suitable for abundance determinations because of (1) their visibility
over a wide $T_{\rm eff}$ range, (2) no concern for any strong non-LTE 
effect, and (3) being presumably insensitive to chromospheric activity
because of their deep-forming nature. We, therefore, decided to
conduct an extensive oxygen abundance study for a number of early F to
late G stars of this cluster by using these O~{\sc i} 6156--8 lines.\\ 
---Do [O/H] values of Hyades F--G stars are sufficiently uniform? 
Or, alternatively, do they show any trend as in the case of A stars?\\
---How are they compared with other metals? A supersolar tendency
is also observed?

\subsection{Carbon}

Regarding carbon, available abundance studies for Hyades stars 
are rather insufficient compared to the case of oxygen.

First, C abundances of Hyades A-type stars 
(9000~K~$\gtsim T_{\rm eff} \gtsim 7500$~K) were derived by 
Varenne and Monier (1999) as well as Gebran et al. (2010) 
based on C~{\sc i} lines such as the one at 5052.17~$\rm\AA$, 
and they obtained almost the same results: [C/H] shows 
a diversity between $\sim -0.8$ and $\sim +0.1$ (where Am stars 
tend to show particularly large deficiencies), though its 
dependence on $v_{\rm e}\sin $ is not clear as is the case for [O/H]. 

Then, as to the [C/H] values of Hyades F-stars ($T_{\rm eff} \ltsim$~7000~K)
we know only a few published studies.\\
--- Tomkin and Lambert (1978) derived [C/H] values for two 
Hyades F-type stars (45 Tau = HD~26462 and HD~27561) by 
using C~{\sc i} lines in the red and near-IR region (such as those
at 7111--9~$\rm\AA$) and obtained slightly supersolar values of 
[C/H] = +0.06 and +0.18, respectively.\\
---Friel and Boesgaard (1990) carried out C-abundance determinations
for 13 Hyades F-stars of $T_{\rm eff} \sim$~6000--7000~K by using
the C~{\sc i} 6587, 7110, 7111--9 lines and derived the marginally
supersolar result on the average ($\langle$[C/H]$\rangle \simeq +0.04$; 
with a standard deviation of 0.07), which means [C/Fe] $\simeq 0.0$
as Fe also shows a slight excess of this amount.\\ 
---Varenne and Monier (1999) as well as Gebran et al. (2010) 
(mentioned above) concluded that carbon in Hyades F-type stars is 
essentially homogeneous and near solar ([C/H]~$\sim 0$) with only 
a small scatter of $\ltsim$~0.1--0.2~dex.

Recently, the abundance of carbon (especially in relation to oxygen) 
in solar-type stars has acquired growing astrophysical interest
among astronomers, given its important role played in the chemical 
evolution of the Galaxy. Takeda and Honda (2005) showed in their 
study of 160 FGK stars that [C/O] ratio is supersolar ($>0$) at 
the metal-rich regime ([Fe/H]~$>0$) with its extent progressively 
increasing toward a higher [Fe/H], which was first pointed out 
by Gustafsson et al. (1999), because the decreasing rate of [C/H] 
with [Fe/H] is slower than that of [O/H]. \footnote{The [C/O] ratio 
may also be an important key for spectroscopically sorting out 
planet-host stars, as recently claimed by Petigura and Marcy (2011).}
It is thus interesting to check whether the [C/O] ratio of Hyades stars
([Fe/H]~$\sim$0.1--0.2) is supersolar as in nearby field FGK 
stars of [Fe/H]~$>0$. This will provide us with important 
information for understanding the chemical composition of the 
primordial gas, from which cluster stars were formed. 

Accordingly, it makes our alternative aim of this study to establish 
the abundances of carbon for Hyades F--G stars as precisely as possible,
in order to examine the degree of homogeneity (how large is the 
star-to-star scatter of [C/H]?) and the abundance ratios
relative to the Sun (how much are the values of [C/H] and [C/O]
on the average?).
For this purpose, we apply spectrum-fitting to C~{\sc i} lines at 
7111--7119~$\rm\AA$ (also used by Friel and Boesgaard 1990), 
where several C~{\sc i} lines of appreciable strengths are confined
and reliable C-abundance determinations may be expected.

\subsection{Lithium}

Besides, this study also focuses on lithium of Hyades F--G stars,
because we are particularly interested in this element in connection
with our recent work and the Li~{\sc i} 6708 line is measurable
in our spectra.

Actually, a number of investigations have been published so far 
on the Li abundances of Hyades late-A through early-K dwarfs
(Herbig 1965; Wallerstein et al. 1965; Cayrel et al. 1984; 
Boesgaard \& Tripicco 1986; Boesgaard \& Budge 1988; 
Burkhart \& Coupry 1989, 2000;  Soderblom et al. 1990; 
Thorburn et al. 1993), and the qualitative trend of $A$(Li) 
is quite well established; that is, a conspicuous Li chasm in early-F stars 
around $T_{\rm eff} \sim$~6300--6800~K, a progressive decline of $A$(Li) 
with a decrease in $T_{\rm eff}$ for G stars at $T_{\rm eff} \ltsim 6000$~K. 

Yet, we point out that most of these studies tend to place emphasis
on stars of specific spectral types (e,g., either A/Am stars, F stars, 
or G stars), and are based on classical-type analysis using equivalent 
widths ($EW$). It would be worthwhile to revisit the Li abundances of
early-F through late-G stars covering a wider $T_{\rm eff}$ range 
by applying the spectrum-synthesis technique to the Li~{\sc i} 6708 line 
and taking into account the non-LTE effect, in a consistent manner such
as done by Takeda and Kawanomoto (2005), which may help to clarify 
their trend as well as dependence (if any) on stellar parameters 
to a quantitatively higher precision. Specifically, we would like 
to find answers to the following questions, for example, resulting 
from our recent related work:\\
---Do the $A$(Li) values of Hyades F stars connect well with those of 
A-type stars which we recently determined (Takeda et al. 2012)?\\ 
---Is the $A$(Li) vs. $v_{\rm e}\sin i$ relation established for 
field solar-analog stars (Takeda et al. 2007, 2010) also observed 
in Hyades early-G stars?

\subsection{Construction of This Paper}

The remainder of this paper is organized as follows.
The adopted observational data for 68 Hyades F--G stars are 
explained in section 2, while the assigned atmospheric parameters 
and model atmospheres are mentioned in sections 3 and 4,
respectively. Section 5 describes 
the procedures of our abundance determinations, which are made up 
of spectrum-synthesis fitting for finding the best-fit solutions
and inverse evaluations of equivalent widths (from which changes 
to parameter perturbations are estimated). The finally resulting 
abundances of Li, C, and O (along with Fe) and their trends are 
discussed in section 6. The conclusions are summarized in section 7. 

\section{Observational Data}

The targets of this study are 68 main-sequence stars of F--G spectral 
type (corresponding to $T_{\rm eff} \sim$~5000--7000~K) 
belonging to the Hyades cluster, which were selected from 
de Bruijne, Hoogerwerf, and de Zeeuw's (2001) catalogue 
(``tablea1.dat'' therein), as given in table 1. 
These program stars are plotted on the color--magnitude diagram 
in figure 1a, based on the data of $B-V$ and $M_{V}$ taken (or computed
with the help of the parallax) from Hipparcos catalogue (ESA 1997). 

\setcounter{figure}{0}
\begin{figure}
  \begin{center}
    \FigureFile(60mm,120mm){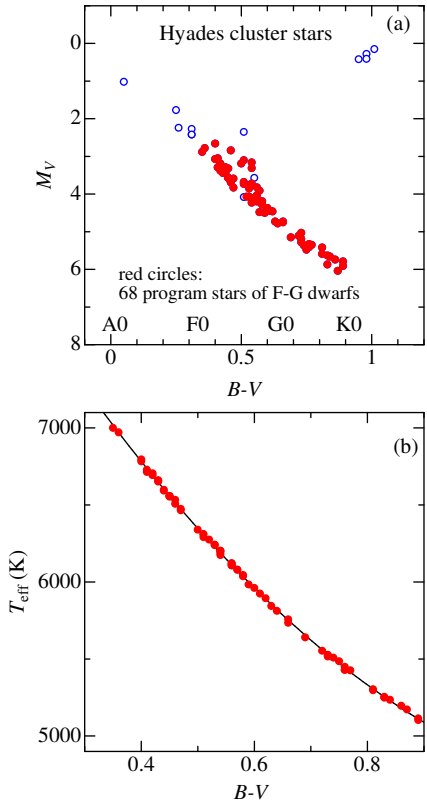}
  \end{center}
\caption{
(a) Hyades cluster stars plotted on the color ($B-V$) vs. magnitude 
($M_{V}$: absolute visual magnitude) diagram. The 68 targets in the 
present study are expressed in (red) filled symbols, while 
other important Hyades stars (mostly A-type dwarfs and G--K giants) 
which are not included in our sample are also shown for comparison
in open (blue) symbols. (b) Correlation plots between the adopted 
$T_{\rm eff}$ and $B-V$ color. The quadratic relation derived from 
the least-squares fit, $T_{\rm eff} = 9000 - 6503 (B-V) + 2397 (B-V)^2$,
is also depicted by the solid line.
}
\end{figure}

The observations were carried out on 2003 December 9--16 and
2004 March 1--4 by using the HIgh-Dispersion Echelle Spectrograph 
(HIDES; Izumiura 1999) at the coud\'{e} focus of the 188~cm reflector 
of Okayama Astrophysical Observatory. Equipped with a 
4K$\times$2K CCD detector\footnote{This was the situation (only one CCD) 
at the time of observations in 2003--2004. At present, HIDES has 
three mosaicked 4K$\times$2K CCDs with the whole wavelength coverage 
of $\sim 3700$~$\rm\AA$.} at the camera focus, the HIDES 
spectrograph enabled us to obtain an echellogram covering a 
wavelength range of 5950--7170~$\rm\AA$ with a resolving power 
of $R \sim 67000$ (case for the normal slit width of 200~$\mu$m) 
in the mode of red cross-disperser.

The reduction of the spectra (bias subtraction, flat-fielding, 
scattered-light subtraction, spectrum extraction, wavelength 
calibration, and continuum normalization) was performed by using 
the ``echelle'' package of the software IRAF\footnote{
IRAF is distributed by the National Optical Astronomy Observatories,
which is operated by the Association of Universities for Research
in Astronomy, Inc. under cooperative agreement with the National 
Science Foundation.} in a standard manner. 
For most of the targets, we could accomplish sufficiently high S/N 
ratio of $\sim$~200--300. 

Besides, since the region comprising C~{\sc i} 7111--9 lines
(which we used for C-abundance determination)
is partly contaminated by telluric water vapor lines,
we removed them by dividing the raw spectrum
of each star by a relevant spectrum of $\alpha$~Leo (rapid rotator) 
by using the IRAF task {\tt telluric}. 
A demonstrative example of this elimination process is depicted 
in figure 2. Actually, the telluric features could be satisfactorily 
cleared away by this procedure for all the 68 targets.

\setcounter{figure}{1}
\begin{figure}
  \begin{center}
    \FigureFile(60mm,60mm){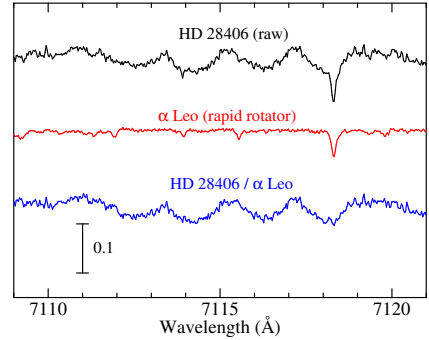}
  \end{center}
\caption{
Example of how the telluric lines (due to H$_{2}$O vapor) are 
removed in the 7109--7121~$\rm\AA$ region comprising C~{\sc i} lines, 
shown for the representative case of HD~28406. 
Dividing the raw stellar spectrum (upper, black) by the spectrum 
of a rapid rotator $\alpha$ Leo (middle, red) results in the final 
spectrum (bottom, blue).
}
\end{figure}

\section{Atmospheric Parameters}

Regarding the effective temperature ($T_{\rm eff}$) and 
surface gravity ($\log g$) for each of the 68 target stars,
we adopted the values directly evaluated from mass, radius, and 
luminosity by de Bruijne, Hoogerwerf, and de Zeeuw (2001) and 
given in their ``tablea1.dat''. This is due to our policy
of making our analysis as consistently as possible for all the
program stars covering a rather wide $T_{\rm eff}$ range,
since widely used spectroscopic determinations of $T_{\rm eff}$ and 
$\log g$ using F~{\sc i} and Fe~{\sc ii} lines can not be applied 
to F-type stars of larger rotational velocity (owing to the
difficulties in equivalent-width measurements) even if applicable
to sharp-lined (mostly G-type) stars. These $T_{\rm eff}$ and 
$\log g$ values are given in table 1, as well as in the 
electronic table E1 (tableE1.dat). The resulting $T_{\rm eff}$ vs. 
$B-V$ relation is displayed in figure 1b, where we can recognize
a tight relationship between these two quantities.

As for the assignment of the microturbulence ($\xi$) to each star, 
we invoked the following empirical relations,\footnote{These empirical
approximations for $\xi$ applicable to FGK stars were 
already reported in Takeda (2008; cf. page 314 therein). 
Note, however, the inequality
signs discriminating two $T_{\rm eff}$ ranges are erroneously
reversed in that article. Given here are the correct ones.}
which were established from the linear-regression analysis on 
the $\xi$ results of 160 FGK stars determined by Takeda et al. (2005).
\begin{equation}
\xi  = 9.9 \times 
10^{-4} T_{\rm eff}  - 0.41 \log g  - 2.92  
\end{equation}
(for $T_{\rm eff} > 5800~{\rm K}$)
\begin{equation}
\xi  = 5.6 \times 
10^{-4} T_{\rm eff}  - 0.31 \log g  - 0.79
\end{equation}
(for $T_{\rm eff} < 5800~{\rm K}$), \\
where $\xi$, $T_{\rm eff}$, $g$
are in the units of km~s$^{-1}$, K, and cm~s$^{-2}$, respectively.
How well these equations (1) and (2) approximate the $\xi$ results
of Takeda et al. (2005) is demonstrated in figures 3a and 3b,
where we can see that these formulae can reproduce the real data
within a few tenths km~s$^{-1}$ for most cases.
Thus, the $\xi$ values for the 68 program stars were computed from
$T_{\rm eff}$ and $\log g$ by using these relations, as presented
in table 1.

It is interesting to compare such assigned values of $T_{\rm eff}$, 
$\log g$, and $\xi$ (which we call as ``standard parameters'') 
with those spectroscopically determined based on the conventional
method using Fe~{\sc i} and Fe~{\sc ii} lines. As an example,
such a comparison with those of Paulson et al.'s (2003)
spectroscopic parameters for Hyades G-type stars (40 stars in common
with our sample) is displayed in figure 4a--4c.
In addition, we also tried establishing spectroscopic parameters 
by ourselves for selected 37 comparatively sharp-lined G dwarfs 
(out of total 68 samples) based on the equivalent widths of 
$\sim 100$ Fe~{\sc i} lines and $\sim 10$ Fe~{\sc ii} lines 
measured on the same spectra as used in this study, and the results
were briefly reported in Takeda (2008). So, we here compare them 
with the standard parameters in figures 5a--5c,
while presenting the detailed data of these spectroscopic parameters 
in electronic table~E2 (tableE2.dat).
We note the following characteristics from these figures.
\begin{itemize}
\item Spectroscopically determined $T_{\rm eff}$ tends to be 
systematically higher by the adopted $T_{\rm eff}$ 
by $\sim 100$~K (figures 4a and 5a).
\item Some spectroscopically determined $\log g$ are
appreciably lower (by $\sim$~0.2--0.3~dex) than the adopted 
$\log g$, though overall agreement is not so bad 
within $\sim \pm 0.1$~dex (figures 4b and 5b).
\item Spectroscopically determined $\xi$ tends to be
somewhat lower than our empirical formula values by 
$\ltsim 0.5$~km~s$^{-1}$ in Paulson et al.'s (2003) results
(cf. figure 4c), while this inequality is just reversed in 
our results where spectroscopic $\xi$ values are higher 
by a few tenths km~s$^{-1}$ than the formula values (figure 5c).
\end{itemize}

\setcounter{figure}{2}
\begin{figure}
  \begin{center}
    \FigureFile(60mm,120mm){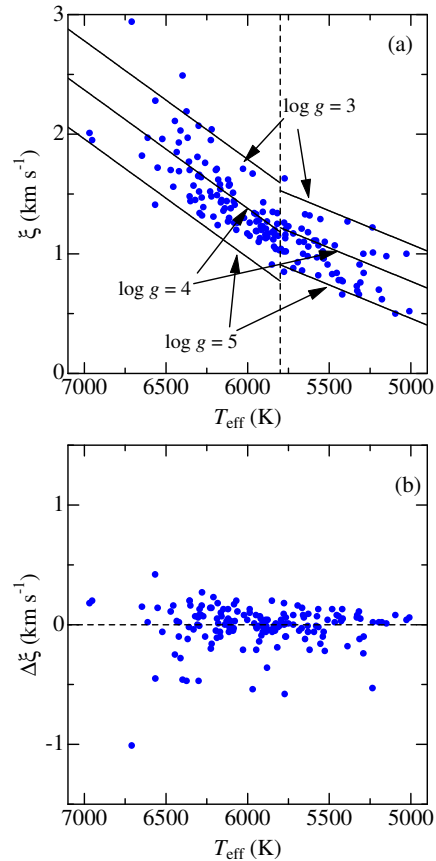}
  \end{center}
\caption{
(a) The $\xi$ vs. $T_{\rm eff}$ plots based on the data of 160 F, G, and K 
stars derived by Takeda et al. (2005). The approximate relations given by
equations (1) and (2) corresponding to $\log g$ = 3, 4, and 5 are shown by 
solid lines. (b) The $\Delta \xi$ vs. $T_{\rm eff}$ plots for the
$\xi$ data of 160 stars in panel (a). Here, $\Delta \xi$ is the residual 
defined as $\xi_{\rm formula} - \xi_{\rm true}$, where $\xi_{\rm formula}$ 
is given by equations (1) and (2) for ($T_{\rm eff}$, $\log g$) of each star.
}
\end{figure}

\setcounter{figure}{3}
\begin{figure}
  \begin{center}
    \FigureFile(80mm,120mm){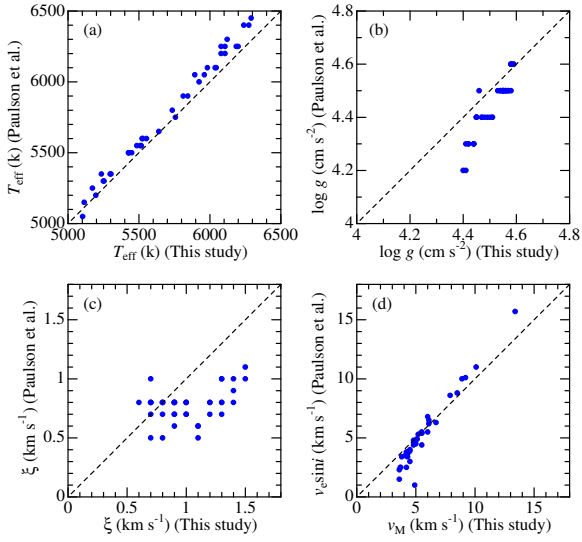}
  \end{center}
\caption{Comparison of our adopted stellar parameters
with those derived by Paulson et al. (2003) (40 stars in common). 
(a) $T_{\rm eff}$,
(b) $\log g$, (c) $\xi$, and (d) $v_{\rm e}\sin i$ (tentatively compared 
with our $v_{\rm M}$, which is a rough measure of $v_{\rm e}\sin i$).
}
\end{figure}

\setcounter{figure}{4}
\begin{figure}
  \begin{center}
    \FigureFile(80mm,120mm){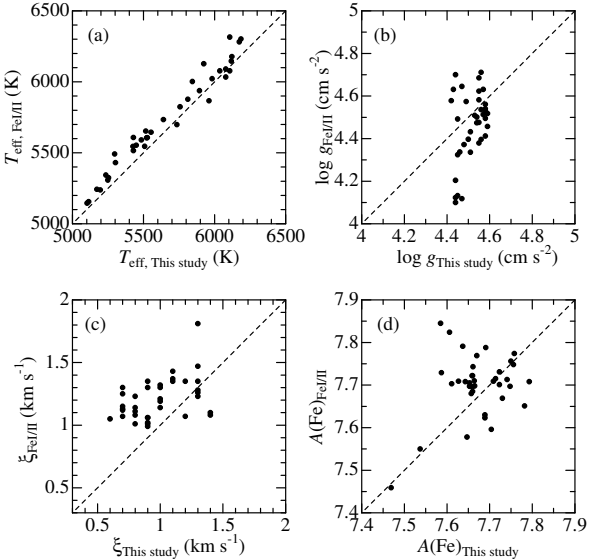}
  \end{center}
\caption{Comparison of our adopted stellar parameters
with the spectroscopically established values by using
Fe~{\sc i} and Fe~{\sc ii} lines for 37 G-type stars 
in common (cf. Takeda 2008). (a) $T_{\rm eff}$,
(b) $\log g$, (c) $\xi$, and (d) $A$(Fe).
}
\end{figure}

\section{Model Atmospheres}

The model atmosphere for each star was then constructed
by two-dimensionally interpolating Kurucz's (1993) ATLAS9 
model grid in terms of $T_{\rm eff}$ and $\log g$, where
we applied the solar-metallicity models computed with 
a microturbulent velocity of 2~km~s$^{-1}$ (``ap00k2.dat''). 

These original ATLAS9 models approximately include the convective 
overshooting effect in an attempt to simulate the real convection 
as possible. 
It has been occasionally argued, however, that this treatment
may cause inconsistencies with observational quantities (e.g., 
colors or Balmer line profiles) and even the classical
pure mixing-length treatment ``without overshooting'' would be 
a better choice (e.g., Castelli et al. 1997). Since lines tend to 
become somewhat weaker in ``with overshooting'' atmospheres as 
compared to ``without overshooting'' cases because of the lessened 
temperature gradient in the lower part of the atmosphere,  some 
difference may be expected in resulting abundances between these 
two cases, especially for comparative higher $T_{\rm eff}$ stars
(i.e., early G to late A; cf. Fig. 24 of Castelli et al. 1997) 
where the convection zone due to hydrogen ionization comes close 
to the bottom of the atmosphere.

In order to maintain consistency with our previous work,
we adopt the original ATLAS9 models ``with overshooting''
as the standard models throughout this study. However,
we also tried deriving abundances in our spectrum-fitting analysis 
by applying ``no overshooting'' models\footnote{
Available from Kurucz's web site 
$\langle$http://kurucz.harvard.edu/$\rangle$ as
``ap00k2nover.dat''.} (as well as ``with overshooting'' models) 
and obtained the corresponding abundance changes, 
$\delta^{\rm nover} \equiv A$(no-overshoot)$-A$(with-overshoot),
in order to see whether and how the difference in this treatment
may cause any appreciable effect.

\section{Procedures of Analysis}

\subsection{Synthetic Spectrum Fitting}

Abundance determinations using our spectral-synthesis code,
which is originally based on Kurucz's (1993) WIDTH9 program, 
were carried out by applying the best-fit solution search algorithm 
(Takeda 1995), while simultaneously varying the abundances of 
several key elements ($A_{1}$, $A_{2}$, $\ldots$), macrobroadening 
parameter ($v_{\rm M}$), and the radial-velocity (wavelength) shift
($\Delta \lambda$).

The macrobroadening parameter ($v_{\rm M}$) represents the combined
effects of instrumental broadening, macroturbulence, and rotational 
velocity. As to the form of macrobroadening function, $M(v)$, 
we applied either one of the following two functions 
(rotational-broadening function [for the uniform-disk case]
and Gaussian function), depending on the appearance of 
spectral-line shapes judged by eye-inspection:
\begin{equation}
M(v) \propto \sqrt{1-(v/v_{\rm M})^{2}} 
\end{equation}
\begin{equation}
M(v) \propto \exp[-(v/v_{\rm M})^{2}].
\end{equation}

Specifically, our spectrum fitting was conducted for the following four 
wavelength regions, where the elements whose abundances were treated
as variables are enumerated in each bracket: 
\begin{enumerate}
\item 6080--6089~$\rm\AA$ (Si, Ti, V, Fe, Co, Ni) [primarily for
determinations of $v_{\rm M}$ and Fe abundance] 
\item 6703\footnote{This lower limit was raised up to $\sim 6707$~$\rm\AA$
for G stars ($T_{\rm eff} \ltsim 5800$~K) because of the increased
complexity of the spectra.}--6709~$\rm\AA$ (Li, Fe) 
including Li~{\sc i}~6708 lines 
[for Li abundance determination]
\item 7110--7121~$\rm\AA$ (C, Fe, Ni) including C~{\sc i} 7111--9 lines
[for C abundance determination]
\item 6156--6159~$\rm\AA$ (O, Ti, Fe) including O~{\sc i} 6156--8 lines 
[for O abundance determination]
\end{enumerate}
Note that analyses for the 6080--6089~$\rm\AA$ as well as 
6156--6159~$\rm\AA$ regions are the same as in Takeda and Honda (2005).
Similarly, the analysis for the 6707--6709~$\rm\AA$ region is the same
as in Takeda and Kawanomoto (2005).

Regarding the atomic data of spectral lines (wavelengths, excitation
potentials, oscillator strengths, etc.), we basically invoked the 
compilations of Kurucz and Bell (1995). However, pre-adjustments
of several $\log gf$ values were necessary (i.e., use of empirically 
determined solar $gf$ values) in order to accomplish a satisfactory
match between the observed and theoretical spectrum. The finally 
adopted atomic parameters of important spectral lines are presented
in table 2. As for the damping parameters (which are unimportant
in the present case because very strong lines are absent in the
relevant wavelength regions), the data given in Kurucz and Bell (1995) 
were used; if not available therein, we invoked the default 
treatment of Kurucz's (1993) WIDTH9 program. 

Note that we assumed LTE for all lines at this stage of synthetic 
spectrum-fitting and that this analysis was performed not only
with the standard ``convective overshooting'' model but also with 
the ``no-overshooting'' model, in order to check the difference 
($\delta^{\rm nover}$) between these two treatments (cf. section 4).

Although the convergence of the solutions turned out fairly successful  
for most of the cases, we encountered with some cases where 
convergence was poor (e.g., oscillatory) or abundance solution 
of some specific element even became unstable and divergent.
When any abundance parameter could not be established, we fixed
it at the solar value and retried the calculation. 
After the solutions have been established, we checked by eye 
whether the synthetic theoretical spectrum satisfactorily 
matches the observed spectrum. If the convergence of any 
abundance solution was not sufficiently good, or if the consistency
between theoretical and observed spectrum did not appear 
satisfactorily good at the relevant line position, we judged
this abundance solution to be of ``low reliability.''
How the theoretical spectrum for the converged solutions fits well 
with the observed spectrum for each star is displayed in 
figure 6 (6080--6089~$\rm\AA$ fitting), 
figure 7 (6156--6159~$\rm\AA$ fitting), 
figure 8 (6703--6709~$\rm\AA$ fitting), and
figure 9 (7110--7121~$\rm\AA$ fitting).

The solutions for $A^{\rm LTE}$(Fe) and $v_{\rm M}$ from 6080--6089~$\rm\AA$ 
fitting are presented in table~1. While the LTE abundances 
of $A^{\rm LTE}$(O), $A^{\rm LTE}$(Li), and $A^{\rm LTE}$(C) obtained 
by these spectrum-fitting analyses are not explicitly given, 
they are easily derived from the non-LTE abundances ($A^{\rm NLTE}$) 
and the non-LTE corrections ($\Delta^{\rm NLTE}$) (both given 
in electronic table E1) as $A^{\rm NLTE}- \Delta^{\rm NLTE}$.

\subsection{Macrobroadening Parameter and Rotational Velocity}

It would be appropriate here to remark that the solution of the
macrobroadening parameter ($v_{\rm M}$) derived as a by-product of 
spectrum fitting can be a fairly good indicator of projected 
rotational velocity ($v_{\rm e} \sin i$), on the condition that 
$v_{\rm e} \sin i$ is not too small.

If the rotational velocity is large and spectral lines show
rounded shapes, we used equation (3) (rotational broadening
function), and this choice corresponds to the solutions of 
$v_{\rm M} \gtsim 16$~km~s$^{-1}$ (cf. table 1). 
In this case, $v_{\rm M}$ can naturally be regarded as essentially 
equivalent to $v_{\rm e}\sin i$, since the contributions of
instrumental broadening and macroturbulence (both are
on the order of several km~s$^{-1}$) are anyhow negligible
compared to this extent.

Further, we would point out that $v_{\rm M}$ is still a
good approximation of $v_{\rm e}\sin i$ also for the slower rotation 
case where we used equation (4) (Gaussian broadening function).
This is because, if we require that the FWHMs of the rotation 
function ($\propto \sqrt{1-(v/v_{\rm e}\sin i)^{2}}$) and the
Gaussian function ($\propto \exp[-(v/v_{\rm rt})^{2}]$) be equal, 
we obtain the relation $v_{\rm rt} \simeq 0.94 v_{\rm e}\sin i$
(cf. footnote 12 of Takeda et al. 2008), which guarantees a  
practical equality between these two.

So, as far as $v_{\rm e}\sin i$ is not so small compared with
the instrumental width or the macroturbulence width,
$v_{\rm M} \sim v_{\rm e}\sin i$ is not a bad approximation,
irrespective of the adopted broadening functions.
It should be bear in mind, however, that this relation
does not hold any more at $v_{\rm M} \ltsim 5$~km~s$^{-1}$ 
where the contributions of the instrumental width as well as 
the macroturbulence width become progressively important, 
though $v_{\rm M}$ might still be regarded as a ``qualitative 
measure'' of $v_{\rm e} \sin i$ even in such a slow-rotator 
regime. 

In order to demonstrate this fact, we compare our $v_{\rm M}$ 
results with the $v_{\rm e}\sin i$ values determined in a 
more accurate manner by Paulson et al. (2003) in figure 4d, 
where we can recognize a reasonable correlation between these two.
(Note that the $v_{\rm M}$ data shown in this figure are less 
than 15~km~s$^{-1}$, which means that all of them were derived 
by assuming the Gaussian function.)

\subsection{Equivalent Widths and Abundance Uncertainties}

While the synthetic spectrum fitting directly yielded the abundance 
solutions of Li, C, and O (the main purpose of this study), this approach 
is not necessarily suitable when one wants to evaluate the extent of 
non-LTE corrections or to study the abundance sensitivity to changing 
the atmospheric parameters (i.e., it is rather tedious to repeat 
the fitting process again and again for different assumptions or 
different atmospheric parameters).
Therefore, with the help of Kurucz's (1993) WIDTH9 program\footnote{
This WIDTH9 program had been considerably modified by Y. T. in various 
respects; e.g., inclusion of non-LTE effects, treatment of total 
equivalent width for multi-component lines; etc.}, 
we computed the equivalent widths for Li~{\sc i}~6708 ($EW_{6708}$), 
C~{\sc i}~7113 ($EW_{7113}$; the strongest line among 
the C~{\sc i} lines at 7111--9~$\rm\AA$),
and O~{\sc i}~6158  ($EW_{6158}$), 
 ``inversely'' from the abundance solutions 
(resulting from spectrum synthesis) along with the adopted 
atmospheric model/parameters, which are much easier to handle.
Based on such evaluated $EW$ values, 
the non-LTE ($A^{\rm LTE}$) as well as LTE abundances ($A^{\rm NLTE}$) 
were freshly computed to derive the non-LTE correction 
($\Delta [\equiv A^{\rm LTE} - A^{\rm NLTE}$]).
The procedures for non-LTE calculations are described in Takeda
and Kawanomoto (2005) (for Li) as well as Takeda and Honda (2005) 
(for C and O), which should be consulted for the details.
For the case where $A$(Li) could not be determined (which we 
encountered for several F stars at the ``Li-gap''), we first guessed
the upper-limit of $EW_{6708}$ by the formula
\begin{eqnarray}
EW^{\rm UL}_{6708} \equiv \sqrt{h^{2} + 150^{2}}/(S/N) \;\;\;\; ({\rm m\AA})\\
h \equiv 6708\times1000\times(v_{\rm M}/c) \;\;\;\; ({\rm m\AA}),
\end{eqnarray}
(where 150~m$\rm\AA$ is the approximate intrinsic width defined by
the separation of the components; cf. Takeda \& Kawanomoto 2005),
from which the upper limit of $A$(Li) was derived.

We then estimated the uncertainties in $A$(Li), $A$(C) and 
$A$(O) by repeating the analysis on the $EW$ values while perturbing
the standard atmospheric parameters interchangeably by $\pm 100$~K 
in $T_{\rm eff}$, $\pm 0.1$~dex in $\log g$, and $\pm 0.5$~km~s$^{-1}$ 
in $\xi$ (which are considered to be typical magnitudes of ambiguities; 
cf. section 3). Figures 10 (Li), 11 (C), and 12 (O) graphically show 
the resulting non-LTE abundances ($A^{\rm NLTE}$), equivalent widths 
($EW$), non-LTE corrections ($\Delta^{\rm NLTE}$), 
abundance changes caused by using the no-overshooting model 
($\delta^{\rm nover}$), and abundance variations in response 
to parameter changes ($\delta_{T+}$, $\delta_{g+}$, and $\delta_{\xi+}$), 
as functions of $T_{\rm eff}$. While such obtained non-LTE abundances
of Li, C, and O are given in table 1, the complete results of 
abundances, corrections, and perturbations are presented in electronic 
table E1, where the abundance changes for Fe ($\delta^{\rm nover}$, 
$\delta_{T}$, $\delta_{g}$, and $\delta_{\xi}$, which were 
obtained by repeating the fitting analysis in this case) are also given.
Hereinafter, we often omit the superscript ``NLTE'' of $A^{\rm NLTE}$ 
for denoting the non-LTE abundances of Li, C, and O.

\setcounter{figure}{9}
\begin{figure}
  \begin{center}
    \FigureFile(70mm,100mm){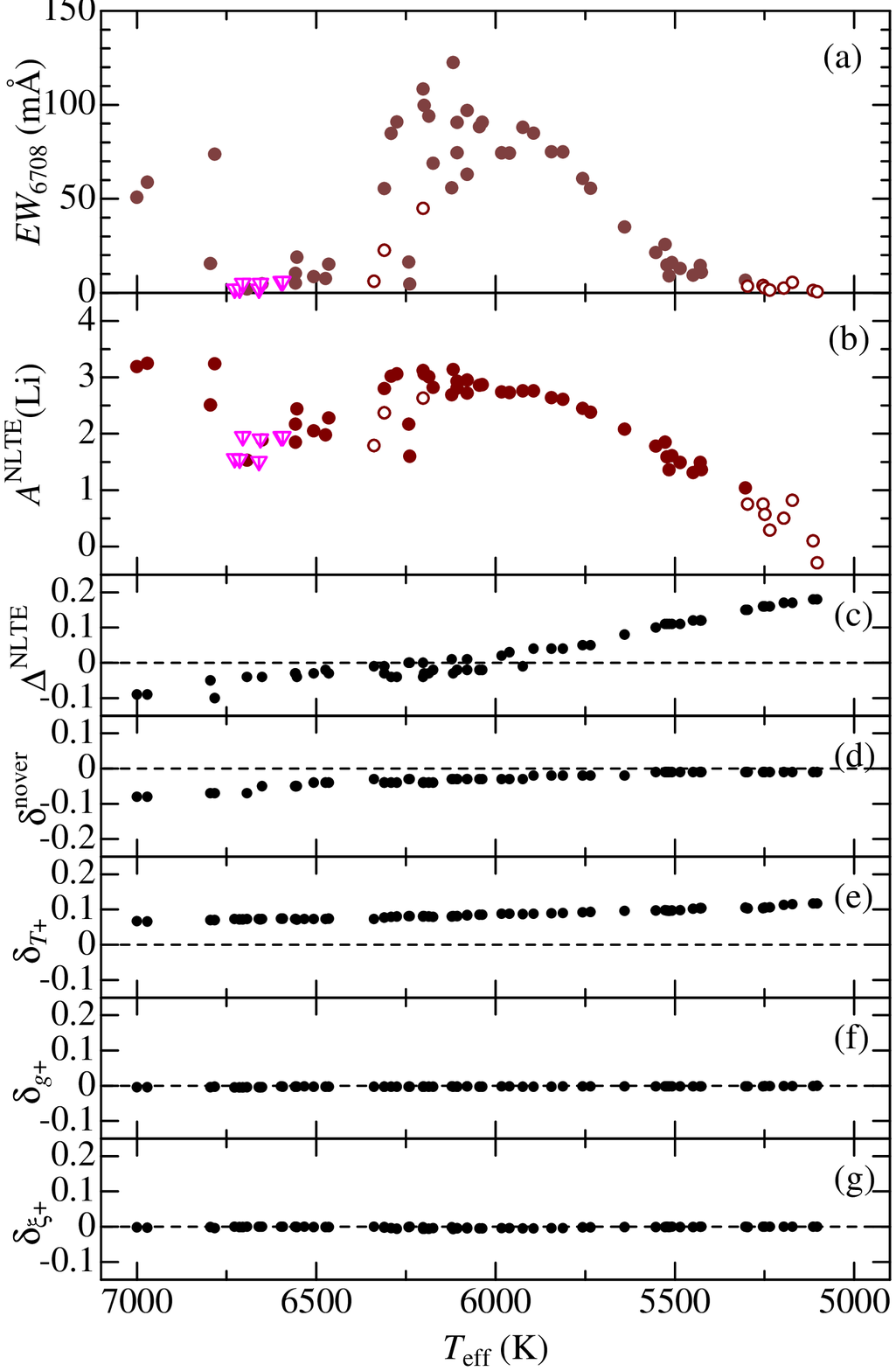}
  \end{center}
\caption{
Li~{\sc i}~6708-related quantities plotted against $T_{\rm eff}$.
(a) $EW$(6708) (equivalent width inversely computed with the 
abundance solution derived from fitting), (b) $A^{\rm NLTE}$(Li)
(non-LTE abundance derived from $EW$), (c) $\Delta^{\rm NLTE}$
(non-LTE correction defined as $A^{\rm NLTE}-A^{\rm LTE}$),
(d) $\delta^{\rm nover}$ (abundance change if no-overshooting model
is used instead of the standard overshooting model), (e)
$\delta_{T+}$ (abundance change in response to an increase of 
$T_{\rm eff}$ by 100~K), (f) $\delta_{g+}$ (abundance change in
response to an increase of $\log g$ by 0.1~dex), and 
(g) $\delta_{\xi +}$ (abundance change in response to an 
increase of $\xi$ by 0.5~km~s$^{-1}$).
In panels (a) and (b), results shown by open symbols
are those with large uncertainties, and downward triangles
(colored in pink) denote upper-limit values.
Note that the ordinate scale of panels (c)--(g) is as $\sim$~6 times
expanded as that of panel (b).
}
\end{figure}

\setcounter{figure}{10}
\begin{figure}
  \begin{center}
    \FigureFile(70mm,100mm){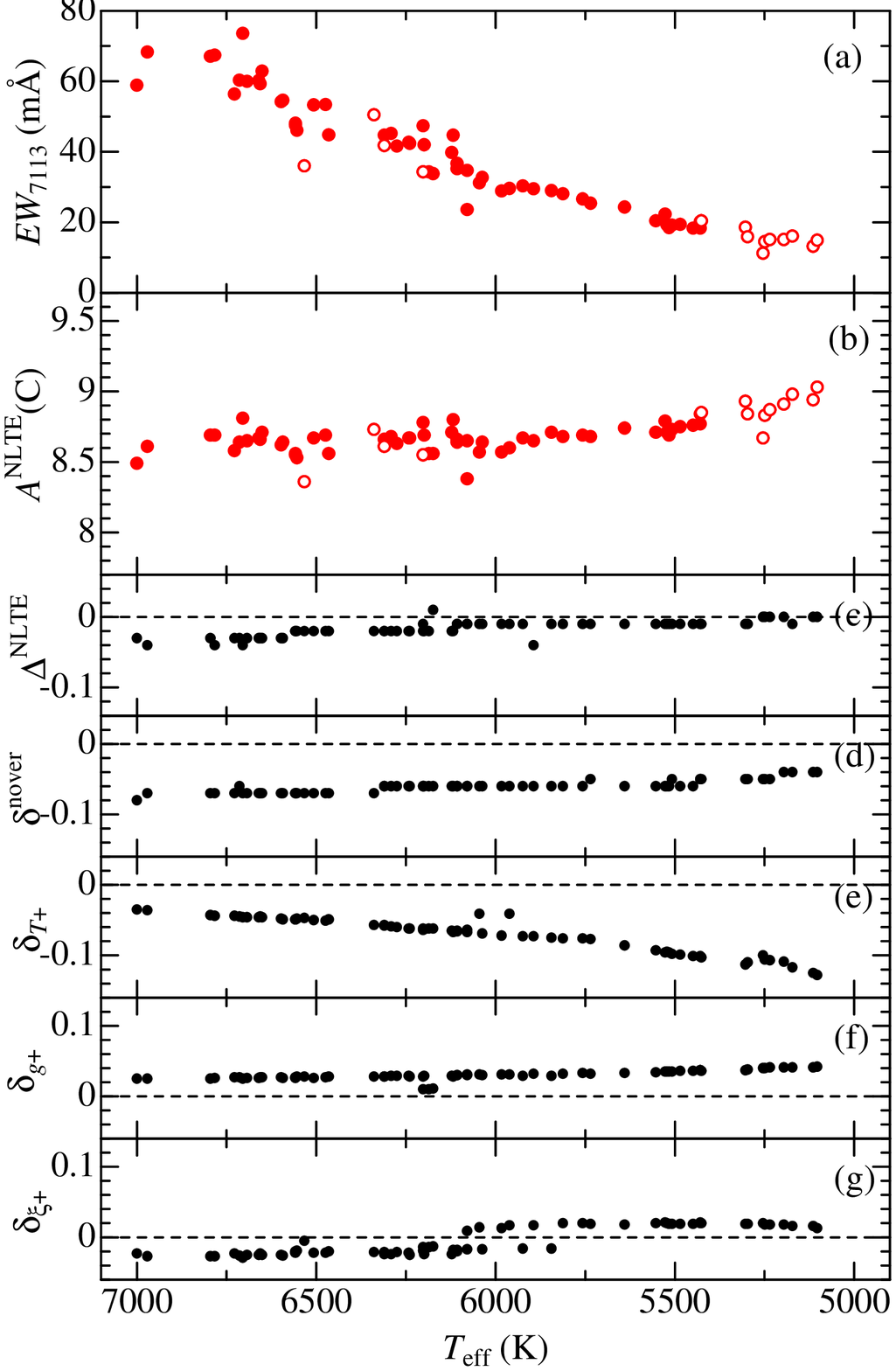}
  \end{center}
\caption{
C~{\sc i}~7113-related quantities plotted against $T_{\rm eff}$.
Note that the ordinate scale of panels (c)--(g) is as $\sim$~7 times 
expanded as that of panel (b).
That the sign of $\delta_{\xi+}$ is positive (which contradicts
the usual trend) at $T_{\rm eff} \ltsim 6000$~K in panel (g)
is interpreted as due to a special effect seen in weak lines
on the linear part of the curve of growth (cf. subsection 3.2
in Takeda 1994). Otherwise, the same as in figure 10.
}
\end{figure}

\setcounter{figure}{11}
\begin{figure}
  \begin{center}
    \FigureFile(70mm,100mm){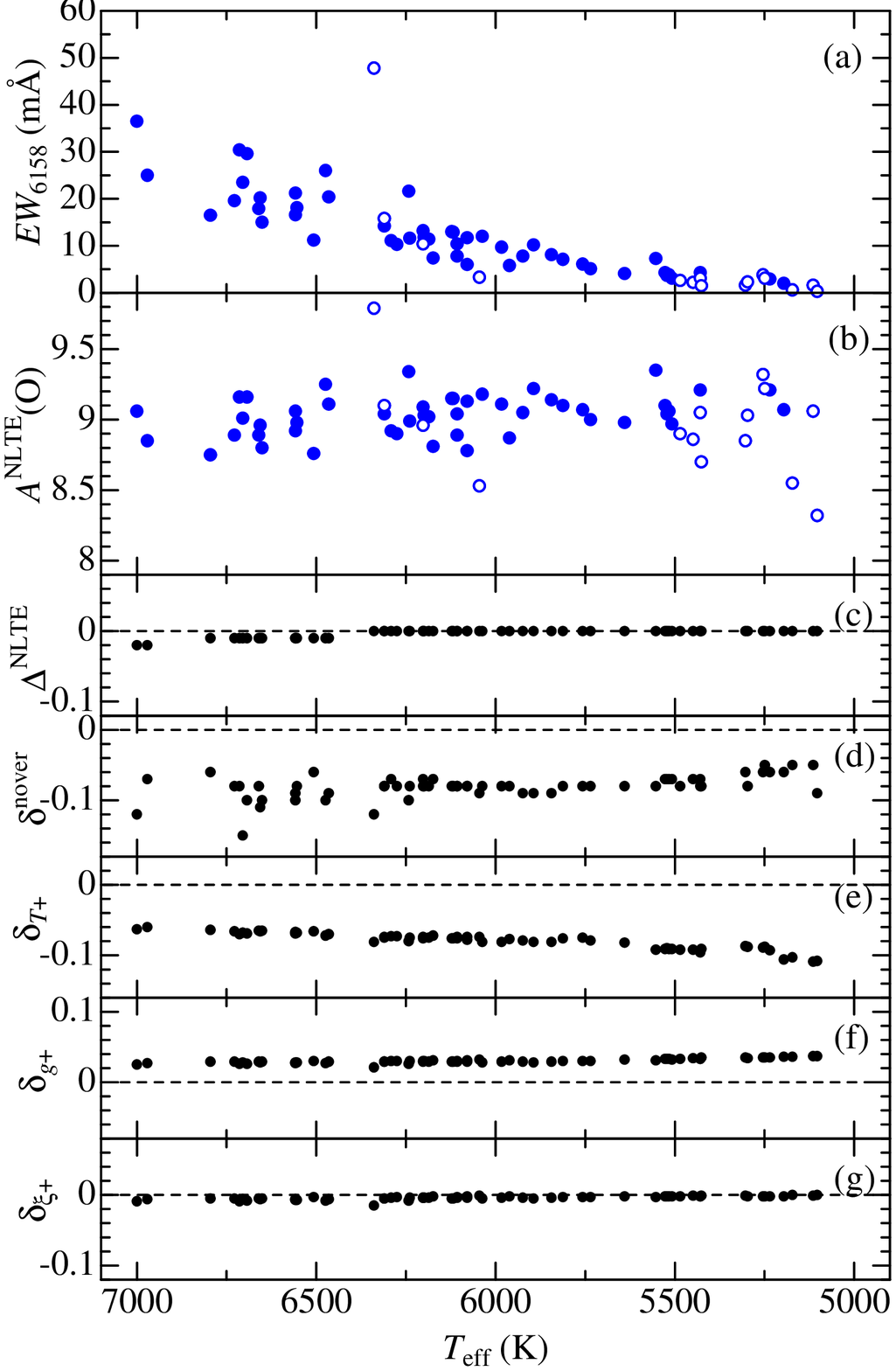}
  \end{center}
\caption{
O~{\sc i}~6158-related quantities plotted against $T_{\rm eff}$.
Note that the ordinate scale of panels (c)--(g) is as $\sim$~7 
times expanded as that of panel (b).
Otherwise, the same as in figure 10.
}
\end{figure}

\section{Discussion}

\subsection{$T_{eff}$-Dependence Problem in A(C), A(O), and A(Fe)}

We first examined whether C and O (elements of our primary concern) 
show essentially the same abundances along the Hyades main sequence. 
A close inspection of figures 11b and 12b revealed that $A$(C)
as well as $A$(O) shows a slightly increasing tendency with a
decrease in $T_{\rm eff}$. Excluding the unreliable determinations
(denoted by open circles), we found from the linear-regression analysis\footnote{
More precisely, our $A$(C), $A$(O), and $A$(Fe) data could be fitted
with the following linear relations in terms of $T_{\rm eff}$
(values in parentheses are the errors of the coefficients):
$A$(C) = $-8.16 (\pm 2.37) \times 10^{-5} T_{\rm eff} + 9.17 (\pm 0.15)$,
$A$(O) = $-1.17 (\pm 0.42) \times 10^{-4} T_{\rm eff} + 9.75 (\pm 0.26)$, and
$A$(Fe) = $-1.00 (\pm 0.14) \times 10^{-4} T_{\rm eff} + 8.24 (\pm 0.09)$.}
$dA({\rm C})/dT_{\rm eff} = -8.2 \times 10^{-5}$ (dex~K$^{-1}$) 
for C (54 stars) and
$dA({\rm O})/dT_{\rm eff} = -1.2 \times 10^{-4}$ (dex~K$^{-1}$)
for O (49 stars), which means a change of $\sim 0.1$~dex over a span of
$\sim 1000$~K.

Interestingly, quite a similar trend is seen in $A$(Fe) given 
in table 1 (64 stars), for which we again found a gradient of 
$dA({\rm Fe})/dT_{\rm eff} = -1.0 \times 10^{-4}$ (dex~K$^{-1}$). 
We note, however, that the situation is not necessarily the same 
for other Fe group elements. 
Figure 13 shows the $A$ vs. $T_{\rm eff}$ relations for six elements 
(Si, Ti, V, Fe, Co, and Ni) derived from the 6080--6089~$\rm\AA$ fitting 
analysis (figure 6). We can recognize from this figure
that any systematic $T_{\rm eff}$-dependence is absent for $A$(Ti) 
and $A$(Co) while $A$(Ni) exhibits a steeper gradient than $A$(Fe).   

\setcounter{figure}{12}
\begin{figure}
  \begin{center}
    \FigureFile(70mm,120mm){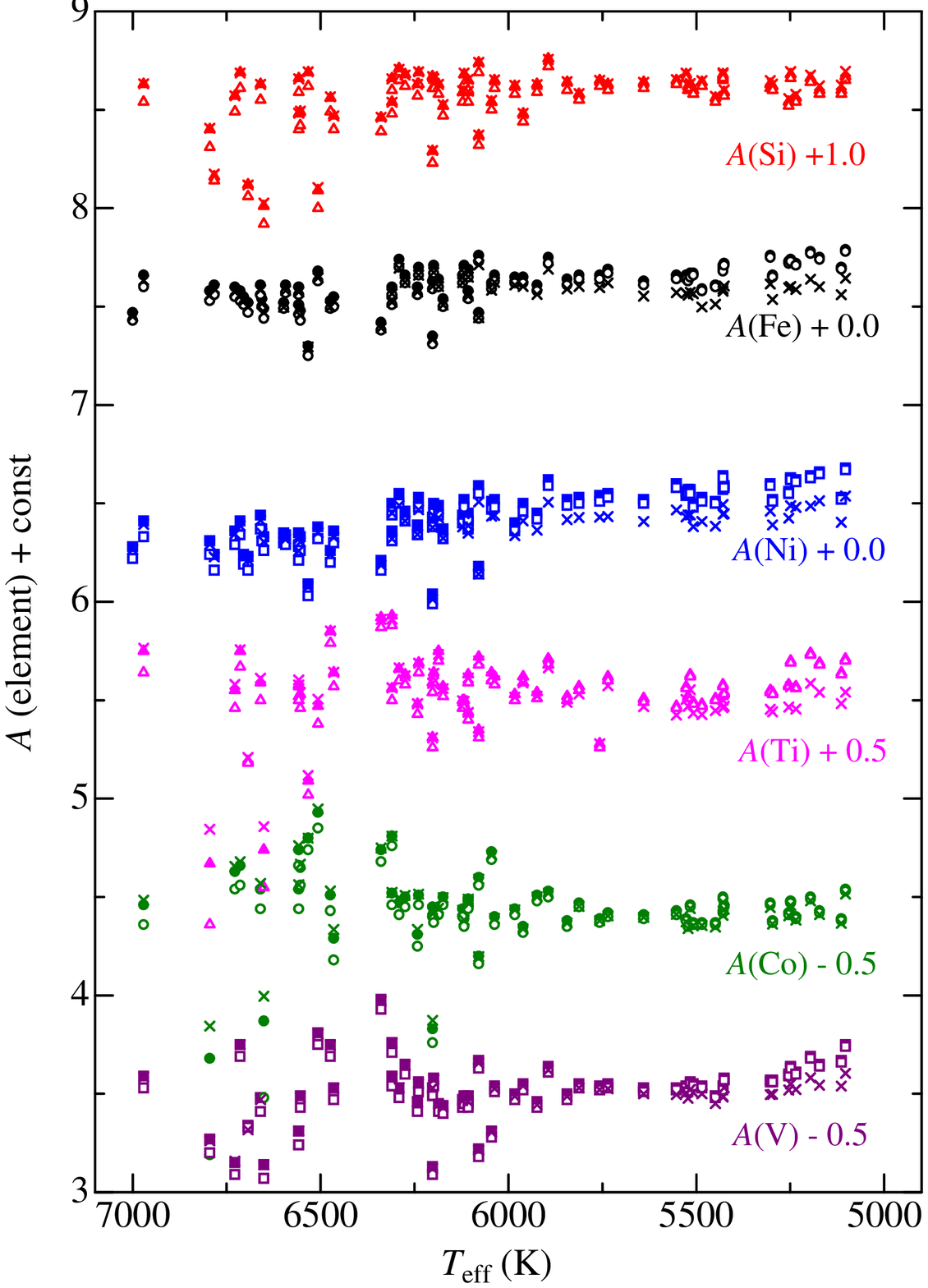}
  \end{center}
\caption{
Abundances of Si, Ti, V, Fe, Co, and Ni (derived from 6080--6089~$\rm\AA$
fitting) plotted against $T_{\rm eff}$. The results obtained
from the ATLAS9 model atmospheres with convective overshooting
(which we adopted as the standard models) are shown by filled symbols,
while those derived with no-overshooting models and those corresponding 
to an increase of microturbulent velocity by 0.5~km~s$^{-1}$ are plotted 
by open symbols and crosses, respectively. 
Vertical offsets of +1.0, +0.5, $-0.5$, and $-0.5$ are applied 
to $A$(Si), $A$(Ti), $A$(Co), and $A$(V), respectively, as described 
in the figure. The results in the $T_{\rm eff}$ range of
6800~K $\gtsim T_{\rm eff} \gtsim 6300$~K (including stars of 
comparatively higher rotation) are subject to larger 
uncertainties (especially for Ti, Co, V;  as recognized by their 
considerable dispersions), and thus should not be taken too seriously.
}
\end{figure}

It would be natural to suspect in the first place that this trend 
may be due to inadequacies in the adopted model atmospheres or 
some improper choice of atmospheric parameters, for which several 
possibilities may be considered:\\
--- The use of ``no-overshooting'' model instead of the standard 
``overshooting'' model can not be the remedy for this trend,
because this acts as a negative correction ($\delta^{\rm nover}<0$)
and its extent $|\delta^{\rm nover}|$ being slightly larger 
toward higher $T_{\rm eff}$ (cf. section 4); i.e., the gradient is 
even more exaggerated (though only marginally) by applying this 
correction (figures 11d, 12d, and 13).\\
--- Meanwhile, the effect of increasing $T_{\rm eff}$ (which is 
probable as spectroscopically determined  $T_{\rm eff}$
turned out to be somewhat larger than the adopted standard $T_{\rm eff}$
by $\sim 100$~K) can cause a $T_{\rm eff}$-dependent correction 
in the direction of suppressing th tendency at least for C and O 
(cf. figures 11e and 12e), though not for Fe. Yet, the extent seems 
still quantitatively insufficient; i.e., even the case of C where 
the largest $T_{\rm eff}$-dependence is observed in $\delta_{T+}$, 
the gradient of $d\delta_{T+}/dT_{\rm eff}$ is as $\sim$~1/2--1/3 
as required to remove the trend.\\
--- Regarding the gravity effect, abundances are practically 
insensitive to a change in $\log g$ (typically a few hundredths dex 
for $\Delta \log g = +0.1$) and this correction hardly depends on 
$T_{\rm eff}$ (figures 11f and 12f).\\
--- As mentioned in section 3, since our spectroscopically determined 
$\xi$ tends to be somewhat larger (by a few tenths km~s$^{-1}$) 
than the adopted $\xi$ based on equations (1) and (2), increasing 
this parameter may be worth consideration. We note that 
the abundances of C and O barely depend on the choice of $\xi$ 
because they are light elements with large thermal velocities 
which makes the contribution of 
non-thermal velocities insignificant (figures 11g and 12g). 
However, the abundance of Fe (along with those of Ti, V, and Ni) 
derived from 6080--6089~$\rm\AA$ fitting is appreciably reduced by 
an increase of $\xi$, and the extent of this downward correction
is larger for lower $T_{\rm eff}$ stars where lines are stronger
and more saturated, which is just in the right direction, as shown
in figure 13 ($\delta_{\xi+}$ for $\Delta\xi = +0.5$~km~s$^{-1}$ is 
$\sim -0.02$~dex, $\sim -0.04$~dex, and $\sim -0.15$~dex at 
$T_{\rm eff} \sim$ 7000~K, 6000~K, and 5000~K, respectively; 
cf. electronic table E1). This could be an explanation (at least 
partly) for the $T_{\rm eff}$-dependence of $A$(Fe), though it is 
not necessarily satisfactory from a quantitative point of view.

Thus, despite these considerations, we could not trace down the reason
for the systematic $T_{\rm eff}$-dependence in $A$(C), $A$(O), and $A$(Fe).
Accordingly, we might as well put the possibility (even if small) 
into our mind that this trend could be real.
In the discussion of the differential abundances relative to the Sun
and their averages over the sample stars (subsection 6.2), we use these 
original abundance results (given in table 1) as they are. Accordingly, 
the existence of such a slight systematic effect should be kept 
in mind; this may cause ambiguities of $\ltsim 0.1$~dex level in 
the averaged abundance depending on which $T_{\rm eff}$ range is used.

\subsection{[C/H], [O/H], and [Fe/H] of Hyades Stars}

We discuss the C, O, and Fe abundances of Hyades F--G stars in comparison
with the solar composition in order to quantitatively establish 
their differential metallicities, with an aim to settle the complicated
situation regarding [C/H] and [O/H] mentioned in subsections 1.2 and 1.3.
As to the reference solar abundances of O and Fe, we adopt 
$A_{\odot}$(O) = 8.81 and $A_{\odot}$(Fe) = 7.53 from Takeda and 
Honda (2005), who derived these values by applying (in exactly the same 
manner as in this study) the 6156--6158~$\rm\AA$ fitting and 
6080--6089~$\rm\AA$ fitting to the moon spectra (cf. section 4 therein).
Meanwhile, the solar carbon abundance was newly determined in this
study by applying the 7110--7121~$\rm\AA$ fitting to the moon spectrum
(taken at this observational period along with other spectra) with the
solar model atmosphere (ATLAS9 model with convective overshooting,
$T_{\rm eff}$ = 5780~K, $\log g = 4.44$, solar metallicity, 
and $\xi = 1$~km~s$^{-1}$), from which we obtained 
$A_{\odot}$(C) = 8.51 ($EW_{7113,\odot} = 21.2$~m$\rm\AA$, 
$\Delta^{\rm NLTE} = -0.01$).

The resulting [Fe/H], [C/H], [O/H], and [C/O] ($\equiv$~[C/H]$-$[O/H]) 
are plotted against $T_{\rm eff}$ in figures 14a--14d, respectively.
Apart from the slight systematic gradient discussed in subsection 6.1,
we can recognize from these figures that the abundances of Fe, C, and O
are reasonably homogeneous with a marginally supersolar tendency.
In the following discussion of the mean abundance and standard deviation,
we exclude the unreliable determinations (open circles) and confine 
only to the reliable results (filled circles).

\setcounter{figure}{13}
\begin{figure}
  \begin{center}
    \FigureFile(70mm,120mm){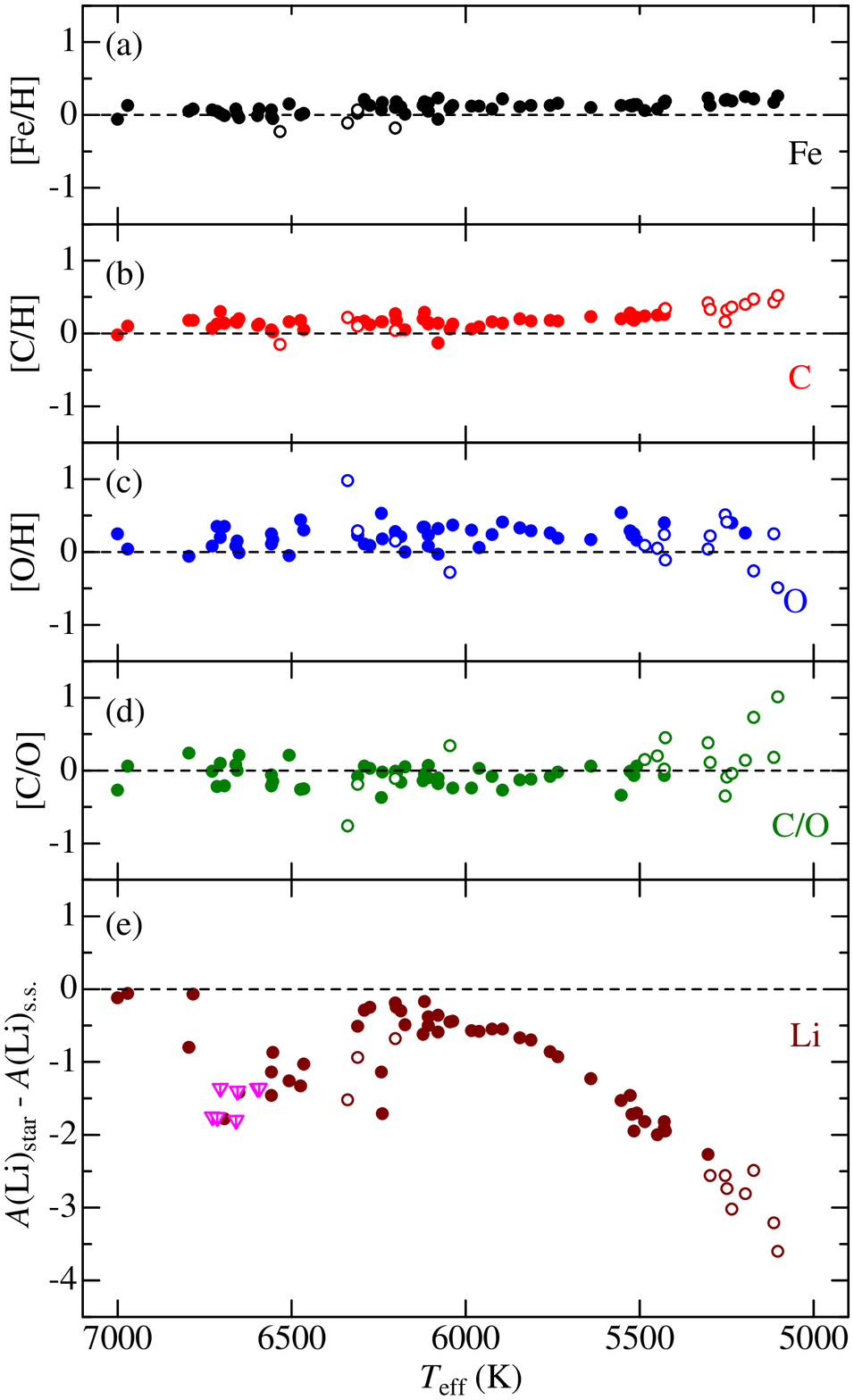}
  \end{center}
\caption{
Logarithmic relative abundances plotted against 
$T_{\rm eff}$. (a) [Fe/H], (b) [C/H], (c) [O/H], (d) [C/O] 
($\equiv$ [C/H]~$-$~[O/H]), and 
$A_{\rm star}$(Li)~$-$~$A_{\rm s.s.}$(Li).
Here, [X/H] is defined as $A_{\rm star}$(X)~$-$~$A_{\odot}$, 
where the adopted reference solar abundances are 
$A_{\odot}$(C) = 8.51, $A_{\odot}$(O) = 8.81, and
$A_{\odot}$(Fe) = 7.53 (cf. subsection 6.2);
while the solar-system abundance of $A_{\rm s.s.}$(Li) = 3.31 is 
used for the case of Li (cf. subsection 6.3).
The meanings of the symbols are the same as in figure 10.
Note that the scale of the ordinate is made to be the same 
for all five panels. 
}
\end{figure}

The mean [Fe/H] (over 64 data) turned out to be 
$\langle$[Fe/H]$\rangle = +0.11$
with the standard deviation ($\sigma$) of 0.08.
So far, a number of published studies on Hyades [Fe/H] values
have yielded results between +0.1 (lower scale) and +0.2 (higher scale)
(see, e.g., figure 32.8 in Takeda 2008). While Takeda (2008) 
derived a higher-scale value of [Fe/H] = +0.19 ($\sigma = 0.05$) 
based on a precise differential study for Hyades early-G stars 
of near-solar $T_{\rm eff}$ ($\sim$~5500--6000~K), our present 
study covering F--G stars implies a result near to the 
lower scale. We consider, however, that this difference may be
due to the existence of a weak $T_{\rm eff}$-dependent gradient
(subsection 6.1), which would make the averaged abundance over
F--G stars slightly lower than that only for G-type stars.

Regarding [C/H] and [O/H], we obtained
$\langle$[C/H]$\rangle$ = +0.15 ($\sigma = 0.08$) from 54 stars
and $\langle$[O/H]$\rangle$ = +0.22 ($\sigma = 0.14$) from 49 stars.
This means that C as well as O are slightly supersolar in Hyades by 
$\sim$~+0.1--0.2~dex just as the case for Fe, and that [C/H] and 
[O/H] are almost uniform over 7000~K~$\gtsim T_{\rm eff} \gtsim 5000$~K 
with only a small dispersion of $\sim \pm 0.1$~dex. 
As to this conclusion of weakly positive nature of [C/H] and [O/H] 
by $\sim 0.2$~dex for this cluster, we can see that most of the 
published values of [C/H] and [O/H] for F--G stars summarized in 
section 1 are more or less consistent with our results, except for 
several studies which suggested near-solar or subsolar C or O  
(e.g., Garc\'{\i}a-Lop\'{e}z et al. 1993 for O; Varenne \& Monnier 
1999 for C; Gebran et al. 2010 for C and O).

Since C and O do not show any sign of deficiency at the $T_{\rm eff}$
range of the ``Li gap'' (6700~K~$\gtsim T_{\rm eff} \gtsim 6300$~K), 
the origin of such a Li trough should be irrelevant to the 
element segregation (atomic diffusion) process, since (if it works) 
light elements such as C, N, O would similarly be influenced.
This confirms the conclusions of Garc\'{\i}a-Lop\'{e}z et al. 
(1993) (for O) and Takeda et al. (1998) (for O and N).

A comparison of $\langle$[C/H]$\rangle$ and $\langle$[O/H]$\rangle$,
indicates that C is slightly less abundant than O.
Actually, the average of [C/O] ratio (for 47 stars) turned out to 
be marginally subsolar as $\langle$[C/O]$\rangle \sim -0.07$ 
($\sigma = 0.14$). It is interesting to note that this
does not conform to the tendency shown by nearby metal-rich stars 
where [C/O] tends to be supersolar in field FGK stars, typically 
by $\sim$~0.1~dex at [Fe/H] $\sim$~0.3 (cf. figure 6d in 
Takeda \& Honda 1995). 
We consider, however, that this is nothing but a natural fluctuation
without any significant meaning. The mean value of [C/O] for slightly 
metal-rich field FGK stars averaged over the metallicity range of 
$0.1 \le$~[Fe/H]~$\le 0.2$ is $\langle$[C/O]$\rangle = +0.026$
with $\sigma = 0.066$ (Takeda \& Honda 2005) and
$\langle$[C/O]$\rangle = +0.025$ with $\sigma = 0.118$ 
(Petigra \& Marcy 2011). This implies that the difference of $\sim 0.1$~dex
($= -0.07-0.03$) from the main trend is still on the order of 1$\sigma$ 
and thus should not be taken seriously.

\subsection{Behavior of Li Abundance}

Finally, we examine the abundances of Li, especially in terms 
of their dependence on $T_{\rm eff}$ and the rotational velocity.  
Our $A$(Li) results and their differences from the solar system 
abundance ($A_{\rm s.s.}$ = 3.31; Anders \& Grevesse 1989) are 
plotted against $T_{\rm eff}$ in figures 10b and 14e, respectively.
We can see from these figures that the well-known characteristics
in the $A$(Li) vs. $T_{\rm eff}$ relation established in previous
studies (see the references cited in subsection 1.4) has been 
firmly corroborated in this study; i.e., an apparent Li chasm
at 6700~K~$\gtsim T_{\rm eff} \gtsim 6300$~K and a progressive
decline of $A$(Li) with a decrease in $T_{\rm eff}$ at
$T_{\rm eff} \ltsim$~6000~K. Although our analysis is different
from the previous work in taking account of the non-LTE corrections
varying from $\sim -0.1$~dex to $\sim +0.2$~dex over the
$T_{\rm eff}$ range of $\sim$~5000--7000~K (cf. figure 10c; 
the difference of the correction sign is because of the fact 
that the dilution of the source function is important at higher 
$T_{\rm eff}$, while the overionization becomes more significant 
at lower $T_{\rm eff}$; cf. section 3 in Takeda \& Kawanomoto 2005), 
these corrections are quantitatively insignificant compared to
the considerably large dynamic range of $A$(Li) amounting up to 
$\sim 3$~dex. Thus, our results superficially look quite similar 
to what has been reported so far.

We note in figure 14e that $A$(Li) at $T_{\rm eff} \gtsim 6800$~K
(on the higher $T_{\rm eff}$ side out of the Li chasm) is almost
the solar-system abundance of $A$(Li)$_{\rm s.s.}$ = 3.31.
Takeda et al. (2012) recently studied the Li abundances of sharp-lined 
A stars including 6 Hyades A/Am stars, among which four stars appear 
to show a weak tendency of decreasing $A$(Li) from $\sim 3.3$ 
($T_{\rm eff} \sim 8000$~K) to $\sim 3.0$ 
($T_{\rm eff} \sim$~7200--7500~K), though Li was 
depleted and unmeasurable in two Am stars. Although we once suspected 
that this might be a continuous extension of the ``Li gap'' to A-stars 
regime ($T_{\rm eff} \gtsim 7000$~K), the present result (preservation
of the primordial Li abundance in stars of 
7000~K~$\gtsim T_{\rm eff} \gtsim 6800$~K) implies that the Li depletion 
mechanism seen in A-type stars is different from the physical process
responsible for the Li gap of Hyades F-type stars.

It was one of our main aims to examine if the Li abundances of Hyades 
stars show any dependence upon the rotational velocity, especially for 
early G-type stars where the evident correlation between $A$(Li) and 
$v_{\rm e}\sin i$ is observed in field solar-analog stars 
(Takeda et al. 2007, 2010).
However, as seen from the tight decline of $A$(Li) from 
$T_{\rm eff} \sim 6000$~K to $\sim 5500$~K (figure 10b)
without showing any considerable scatter seen in field
stars of this $T_{\rm eff}$ range (cf. figure 8 of 
Takeda \& Kawanomoto 2005; figure 9a of Takeda et al. 2007), 
it may be natural to consider that $A$(Li) depends 
only on $T_{\rm eff}$ without any relevance to other parameters,
as least for Hyades G-type stars.
To confirm this, our $A$(Li) results are plotted against $v_{\rm M}$
(measure of $v_{\rm e}\sin i$; cf. subsection 5.2)
in figures 15b (all stars) and 15c (only stars of $T_{\rm eff} < 6000$~K, 
all of which have $v_{\rm M} < 7$~km~s$^{-1}$), from which 
we can read the following characteristics:\\
--- We can not see any significant $v_{\rm M}$-dependence in $A$(Li) of 
F stars ($T_{\rm eff} > 6000$~K) showing a large range of $v_{\rm M}$
($\sim$~10--70~km~s$^{-1}$).\\
--- Regarding G-type stars ($T_{\rm eff} < 6000$~K), we see an 
increasing tendency of $A$(Li) with an increase in $v_{\rm M}$
(figure 15c). We believe, however, that this is nothing but 
an apparent effect due to the $T_{\rm eff}$-dependence of $v_{\rm M}$ 
(i.e., a decrease of $v_{\rm M}$ toward a lower $T_{\rm eff}$; 
cf. figure 15a). Thus, we conclude that the Li abundances of 
Hyades G-type stars are essentially controlled only by $T_{\rm eff}$.
This means that the characteristic $v_{\rm e}\sin i$-dependence
of $A$(Li) observed in field solar-analog stars with ages of
$\sim$~(10--100)$\times 10^{8}$~yr (cf. figure 5g
in Takeda et al. 2010) is absent in younger Hyades stars
(with ages of $\sim 6\times 10^{8}$~yr), which may suggest that such 
a rotation-dependent anomaly is produced during the main-sequence
life time, not in the pre-main-sequence phase.

\setcounter{figure}{14}
\begin{figure}
  \begin{center}
    \FigureFile(70mm,120mm){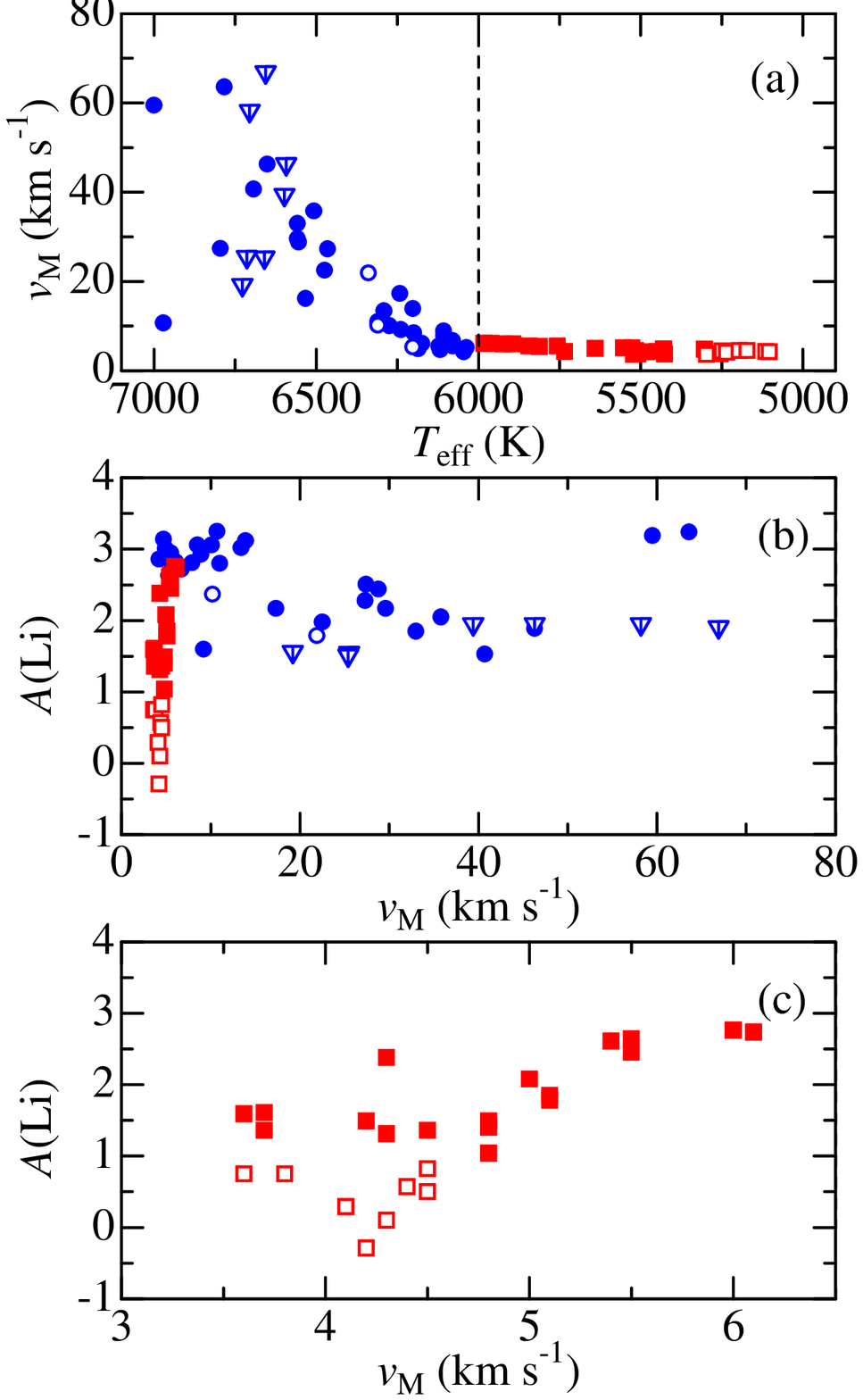}
  \end{center}
\caption{
Relations between $v_{\rm M}$ (macrobroadening velocity derived from 
the 6080--6089~$\rm\AA$ fitting: measure of $v_{\rm e}\sin i$), 
$T_{\rm eff}$, and $A$(Li). (a) $v_{\rm M}$ vs. $T_{\rm eff}$, 
(b) $A$(Li) vs. $v_{\rm M}$ (all data), and (c) $A$(Li) vs. $v_{\rm M}$ 
(only for $T_{\rm eff} < 6000$~K data). Stars for $T_{\rm eff} > 6000$~K
and $T_{\rm eff} < 6000$~K are distinguished by (red) squares
and (blue) circles, respectively. Open symbols and downward triangles
denote that these $A$(Li) results are uncertain values and 
upper-limit values, respectively.
}
\end{figure}

\section{Conclusion}

The C and O abundances of main-sequence stars in the Hyades cluster 
are not yet well established despite their astrophysical importance, 
for which a number of previous studies reported different results. 
Also, the abundances of Li (key element for investigating 
the physical process in the envelope) and Fe (representative of 
metallicity) are worth reinvestigation by taking this opportunity.
Motivated by this situation, we decided to carry out a systematic 
abundance study of these elements for Hyades main-sequence stars 
in the $T_{\rm eff}$ range of $\sim$~5000--7000~K.

Practically, we derived these abundances by applying a spectrum-synthesis 
analysis to four spectral regions at 6080--6089~$\rm\AA$, 6707--6709~$\rm\AA$, 
7110--7121~$\rm\AA$, and 6157--6159~$\rm\AA$ (comprising lines of 
Fe-group elements, Li~{\sc i} 6708 line, C~{\sc i} 7111--7119 lines, 
and O~{\sc i} 6156--8 lines, respectively) based on the 
high-dispersion spectra of 68 selected Hyades F--G type stars 
obtained with the 188~cm reflector and the HIDES spectrograph 
at Okayama Astrophysical Observatory.

It turned out that these C, O, and Fe abundances 
similarly exhibit a marginal $T_{\rm eff}$-dependent gradient 
(i.e., slightly increasing with a decrease in $T_{\rm eff}$; 
typically on the order of $\sim 10^{-4}$~dex~K$^{-1}$) 
Although this might be nothing but an apparent effect due to 
an improper choice of atmospheric parameters, we found it hard 
to give a quantitatively reasonable explanation. 

Apart from this small systematic gradient, the abundances of 
C, O, and Fe in these Hyades stars were found to be fairly uniform
and marginally supersolar with only a small scatter of $\sim 0.1$~dex:
$\langle$[C/H]$\rangle = +0.15$~$(\sigma = 0.08)$,
$\langle$[O/H]$\rangle = +0.22$~$(\sigma = 0.14)$, and
$\langle$[Fe/H]$\rangle = +0.11$~$(\sigma = 0.08)$,
suggesting that the primordial abundances are almost retained.

Regarding Li, we confirmed the well-known $T_{\rm eff}$-dependent 
trend in the Li abundances of Hyades F--G stars reported so far 
(i.e., a conspicuous Li-trough at 
6700~K~$\gtsim T_{\rm eff} \gtsim 6300$~K and a progressive decline
with a decrease in $T_{\rm eff}$ below $\ltsim 6000$~K). 
Since $A$(Li) at 7000~K~$\gtsim T_{\rm eff} \gtsim 6800$~K (a zone
encompassed by the deficiency of Li in A/Am stars and the Li chasm of 
F stars) is almost the solar-system abundance, the Li depletion 
mechanism seen in A-type stars is considered to be different from 
the physical process responsible for the Li gap of Hyades F-type stars.

We concluded that the the surface Li of Hyades stars is essentially 
controlled only by $T_{\rm eff}$ and other parameters such as the 
rotational velocity are almost irrelevant.
A positive correlation between $A$(Li) and stellar rotation, 
which is observed in field solar-analog stars, is not seen in 
these younger early G-type stars of the Hyades cluster. 
This may impose an important constraint on the time scale in 
the build-up of such a rotation-dependent Li anomaly.

\clearpage

\onecolumn

\setcounter{figure}{5}
\begin{figure}
  \begin{center}
    \FigureFile(150mm,200mm){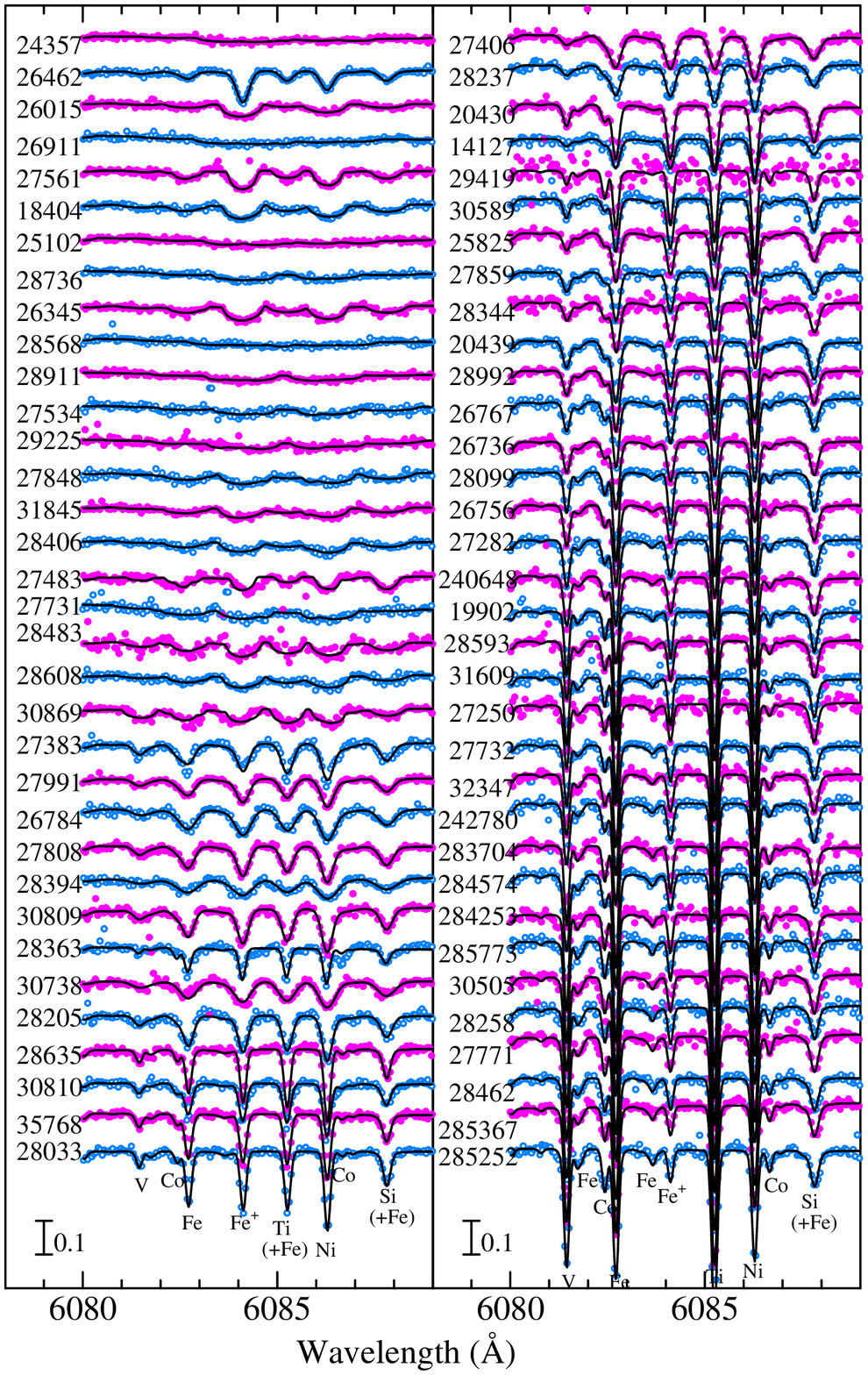}
  \end{center}
\caption{
Synthetic spectrum fitting at the 6080--6089~$\rm\AA$ region 
accomplished by adjusting the macrobroadening velocity ($v_{\rm M}$; 
which is a measure of $v_{\rm e} \sin i$) along with the abundances 
of Si, Ti, V, Fe, Co, and Ni.
The best-fit theoretical spectra are shown by solid lines, 
while the observed data are plotted by symbols.  
In each panel, the spectra are arranged in the descending order 
of $T_{\rm eff}$ as in table 1, and an appropriate offset is 
applied to each spectrum (indicated by the HD number) relative to 
the adjacent one.
}
\end{figure}

\setcounter{figure}{6}
\begin{figure}
  \begin{center}
    \FigureFile(150mm,200mm){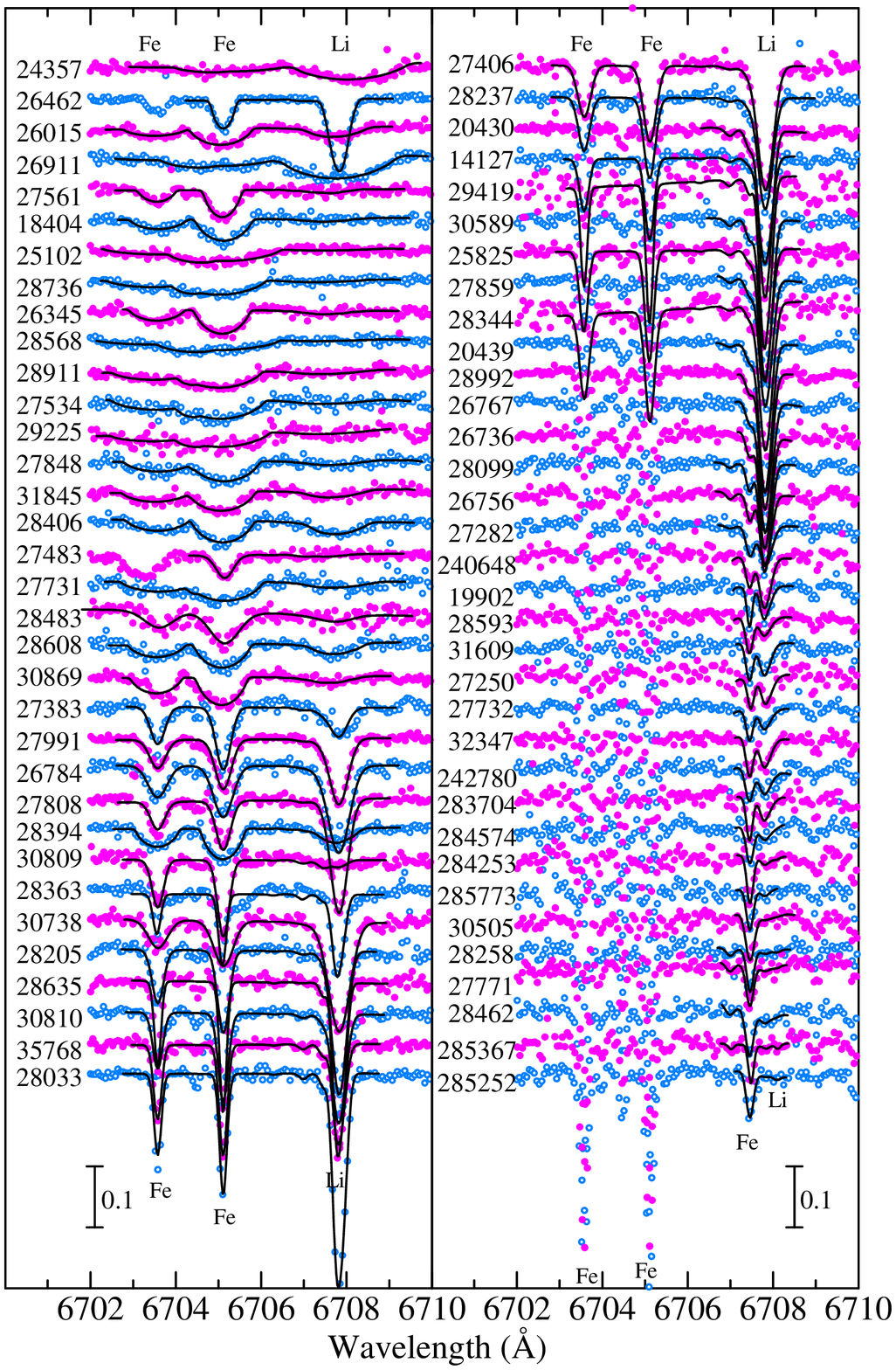}
  \end{center}
\caption{
Synthetic spectrum fitting at the 6703--6709~$\rm\AA$ region 
comprising the Li~{\sc i} 6708 line for determining the 
abundance of Li (and Fe). Otherwise, the same as  in figure 6.
}
\end{figure}

\setcounter{figure}{7}
\begin{figure}
  \begin{center}
    \FigureFile(150mm,200mm){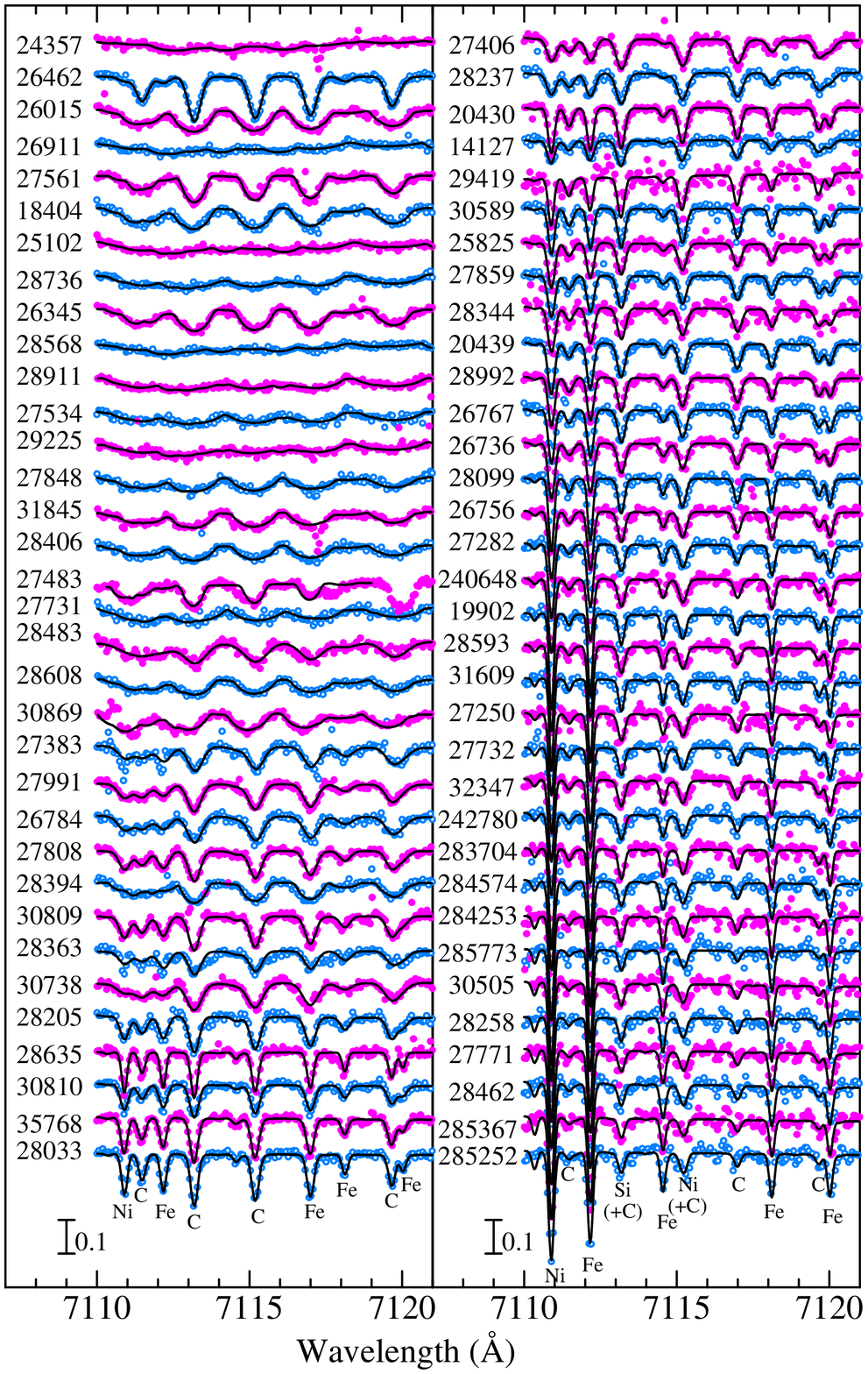}
  \end{center}
\caption{
Synthetic spectrum fitting at the 7110--7121~$\rm\AA$ region 
for determining the abundance of C (along with Fe and Ni). 
Otherwise, the same as in figure 6.
}
\end{figure}

\setcounter{figure}{8}
\begin{figure}
  \begin{center}
    \FigureFile(150mm,200mm){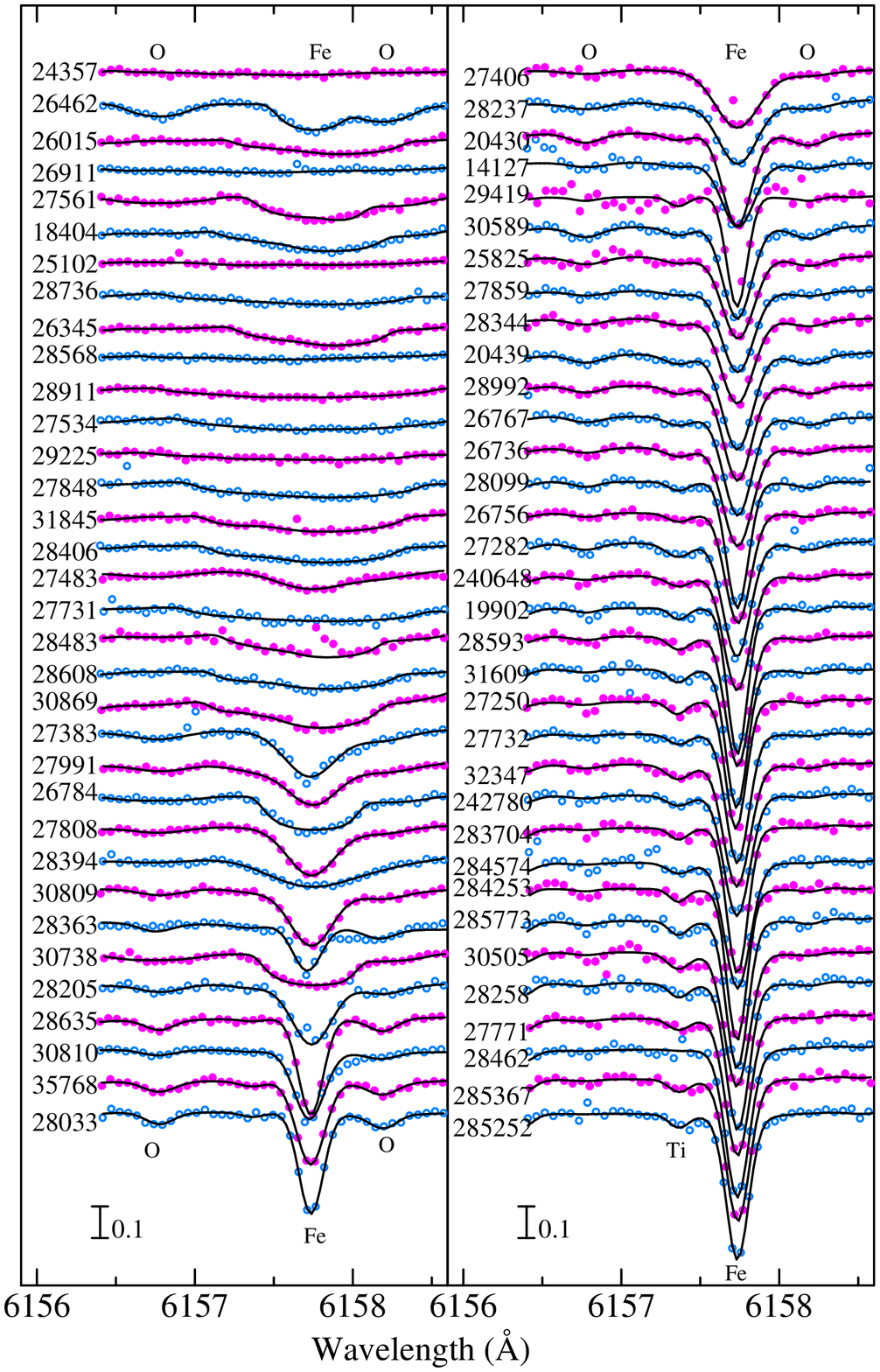}
  \end{center}
\caption{
Synthetic spectrum fitting at the 6156--6159~$\rm\AA$ region 
for determining the abundance of O (along with Fe). 
Otherwise, the same as in figure 6.
}
\end{figure}

\setcounter{table}{0}
\small
\renewcommand{\arraystretch}{0.6}
\setlength{\tabcolsep}{3pt}
\begin{longtable}{ccccccr@{ }ccccc}
\caption{Basic parameters of 68 target stars and the resulting abundances.}
\hline\hline
 HD & $M_{V}$ & $B-V$ & $T_{\rm eff}$ & $\log g$ & $\xi$ & $v_{\rm M}$ & type & $A$(Fe) &
$A$(Li) & $A$(C) & $A$(O) \\
\hline
\endhead
\hline
\endfoot
\hline
\multicolumn{12}{l}{\hbox to 0pt{\parbox{120mm}{\footnotesize
Note.\\
In columns 1 through 6 are presented the HD number, absolute visual magnitude, 
$B-V$ color, effective temperature (in K), logarithmic surface gravity 
(in cm~s$^{-2}$) and microturbulence (in km~s$^{-1}$). Columns 7 nd 8
give the macrobroadening velocity (measure of $v_{\rm e}\sin i$)
and the type of the adopted broadening function (r$\cdots$ rotational function,
g $\cdots$ Gaussian function), respectively. The final abundances of 
$A^{\rm LTE}$(Fe), $A^{\rm NLTE}$(Li), $A^{\rm NLTE}$(C), and 
$A^{\rm NLTE}$(O) are in columns 9--12 (in the usual normalization 
of H = 12.00), where uncertain values are parenthesized and upper-limit 
values are expressed in italic. The stars are arranged in the order 
of descending $T_{\rm eff}$.
}}}
\endlastfoot
\hline
024357 & 2.88 & 0.35 & 7000 & 4.30 & 2.2 & 59.5 & r&  7.47 &   3.19 &   8.49 &   9.06 \\
026462 & 2.78 & 0.36 & 6971 & 4.30 & 2.2 & 10.7 & g&  7.66 &   3.25 &   8.61 &   8.85 \\
026015 & 2.66 & 0.40 & 6795 & 4.32 & 2.0 & 27.4 & r&  7.58 &   2.51 &   8.69 &   8.75 \\
026911 & 3.07 & 0.40 & 6783 & 4.32 & 2.0 & 63.6 & r&  7.61 &   3.24 &   8.69 &  $\cdots$ \\
027561 & 3.05 & 0.41 & 6728 & 4.33 & 2.0 & 19.2 & r&  7.60 &  $<${\it 1.56} &   8.58 &   8.89 \\
018404 & 3.29 & 0.41 & 6714 & 4.33 & 2.0 & 25.5 & r&  7.58 &  $<${\it 1.55} &   8.64 &   9.16 \\
025102 & 3.38 & 0.42 & 6705 & 4.33 & 1.9 & 58.2 & r&  7.55 &  $<${\it 1.95} &   8.81 &   9.01 \\
028736 & 3.19 & 0.42 & 6693 & 4.33 & 1.9 & 40.7 & r&  7.52 &   1.53 &   8.65 &   9.16 \\
026345 & 3.44 & 0.43 & 6660 & 4.34 & 1.9 & 25.4 & r&  7.61 &  $<${\it 1.51} &   8.67 &   8.89 \\
028568 & 3.43 & 0.43 & 6656 & 4.34 & 1.9 & 66.9 & r&  7.54 &  $<${\it 1.91} &   8.66 &   8.96 \\
028911 & 3.41 & 0.43 & 6651 & 4.34 & 1.9 & 46.3 & r&  7.49 &   1.89 &   8.71 &   8.80 \\
027534 & 3.29 & 0.44 & 6598 & 4.34 & 1.8 & 39.4 & r&  7.52 &  $<${\it 1.95} &   8.62 &  $\cdots$ \\
029225 & 3.45 & 0.44 & 6593 & 4.35 & 1.8 & 46.3 & r&  7.61 &  $<${\it 1.95} &   8.64 &  $\cdots$ \\
027848 & 3.32 & 0.45 & 6558 & 4.35 & 1.8 & 33.0 & r&  7.60 &   1.85 &   8.56 &   8.92 \\
031845 & 3.57 & 0.45 & 6558 & 4.35 & 1.8 & 29.6 & r&  7.51 &   2.17 &   8.55 &   9.06 \\
028406 & 3.57 & 0.45 & 6554 & 4.35 & 1.8 & 28.8 & r&  7.48 &   2.44 &   8.53 &   8.98 \\
027483 & 2.84 & 0.46 & 6533 & 4.36 & 1.8 & 16.2 & r&(7.30) &  $<${\it 1.12} & (8.36)&  $\cdots$ \\
027731 & 3.69 & 0.46 & 6507 & 4.36 & 1.7 & 35.8 & r&  7.68 &   2.05 &   8.67 &   8.76 \\
028483 & 3.59 & 0.47 & 6474 & 4.37 & 1.7 & 22.5 & r&  7.53 &   1.98 &   8.69 &   9.25 \\
028608 & 3.83 & 0.47 & 6465 & 4.37 & 1.7 & 27.3 & r&  7.55 &   2.28 &   8.56 &   9.11 \\
030869 & 3.19 & 0.50 & 6339 & 4.39 & 1.6 & 21.9 & r&(7.42) &  (1.79)& (8.73)& (9.79)\\
027383 & 3.68 & 0.51 & 6310 & 4.40 & 1.5 & 10.2 & g&(7.60) &  (2.37)& (8.61)& (9.10)\\
027991 & 3.10 & 0.51 & 6310 & 4.40 & 1.5 & 11.0 & g&  7.55 &   2.80 &   8.66 &   9.04 \\
026784 & 3.73 & 0.51 & 6291 & 4.40 & 1.5 & 13.4 & g&  7.74 &   3.02 &   8.68 &   8.92 \\
027808 & 4.07 & 0.52 & 6275 & 4.41 & 1.5 & 10.1 & g&  7.66 &   3.06 &   8.63 &   8.90 \\
028394 & 3.85 & 0.53 & 6242 & 4.41 & 1.5 & 17.3 & g&  7.60 &   2.17 &   8.67 &   9.34 \\
030809 & 4.07 & 0.53 & 6239 & 4.41 & 1.4 &  9.2 & g&  7.70 &   1.60 &   8.67 &   8.99 \\
028363 & 3.16 & 0.54 & 6202 & 4.42 & 1.4 &  5.3 & g&(7.35) &  (2.63)& (8.55)& (8.96)\\
030738 & 3.72 & 0.54 & 6202 & 4.42 & 1.4 & 13.9 & g&  7.63 &   3.12 &   8.78 &   9.09 \\
028205 & 4.11 & 0.54 & 6199 & 4.42 & 1.4 &  8.5 & g&  7.71 &   3.06 &   8.69 &   9.03 \\
028635 & 4.23 & 0.54 & 6186 & 4.42 & 1.4 &  4.9 & g&  7.64 &   3.01 &   8.56 &   9.02 \\
030810 & 3.31 & 0.54 & 6174 & 4.43 & 1.4 &  6.1 & g&  7.54 &   2.82 &   8.56 &   8.81 \\
035768 & 3.82 & 0.56 & 6122 & 4.44 & 1.3 &  5.5 & g&  7.66 &   2.69 &   8.71 &   9.15 \\
028033 & 4.03 & 0.56 & 6118 & 4.44 & 1.3 &  4.7 & g&  7.71 &   3.14 &   8.80 &   9.15 \\
027406 & 4.20 & 0.56 & 6107 & 4.44 & 1.3 &  8.9 & g&  7.69 &   2.93 &   8.66 &   8.89 \\
028237 & 4.12 & 0.56 & 6107 & 4.44 & 1.3 &  7.9 & g&  7.58 &   2.81 &   8.64 &   9.04 \\
020430 & 3.91 & 0.57 & 6079 & 4.45 & 1.3 &  5.5 & g&  7.76 &   2.95 &   8.65 &   9.13 \\
014127 & 4.48 & 0.57 & 6079 & 4.45 & 1.3 &  6.7 & g&  7.47 &   2.72 &   8.38 &   8.78 \\
029419 & 4.28 & 0.58 & 6045 & 4.45 & 1.2 &  4.2 & g&  7.62 &   2.86 &   8.57 & (8.53)\\
030589 & 4.19 & 0.58 & 6037 & 4.45 & 1.2 &  5.2 & g&  7.66 &   2.87 &   8.64 &   9.18 \\
025825 & 4.50 & 0.59 & 5983 & 4.46 & 1.2 &  6.1 & g&  7.65 &   2.74 &   8.57 &   9.11 \\
027859 & 4.37 & 0.60 & 5961 & 4.47 & 1.1 &  6.1 & g&  7.65 &   2.73 &   8.60 &   8.87 \\
028344 & 4.45 & 0.61 & 5924 & 4.47 & 1.1 &  6.0 & g&  7.61 &   2.76 &   8.67 &   9.05 \\
020439 & 4.46 & 0.62 & 5894 & 4.48 & 1.1 &  6.0 & g&  7.75 &   2.76 &   8.65 &   9.22 \\
028992 & 4.73 & 0.63 & 5844 & 4.49 & 1.0 &  5.5 & g&  7.64 &   2.64 &   8.71 &   9.14 \\
026767 & 4.78 & 0.64 & 5812 & 4.50 & 1.0 &  5.4 & g&  7.66 &   2.61 &   8.68 &   9.10 \\
026736 & 4.73 & 0.66 & 5757 & 4.51 & 1.0 &  5.5 & g&  7.66 &   2.45 &   8.69 &   9.07 \\
028099 & 4.75 & 0.66 & 5735 & 4.51 & 1.0 &  4.3 & g&  7.69 &   2.38 &   8.68 &   9.00 \\
026756 & 5.15 & 0.69 & 5640 & 4.53 & 1.0 &  5.0 & g&  7.63 &   2.08 &   8.74 &   8.98 \\
027282 & 5.11 & 0.72 & 5553 & 4.54 & 0.9 &  5.1 & g&  7.66 &   1.78 &   8.71 &   9.35 \\
240648 & 5.19 & 0.73 & 5527 & 4.54 & 0.9 &  5.1 & g&  7.66 &   1.85 &   8.79 &   9.10 \\
019902 & 5.03 & 0.73 & 5522 & 4.55 & 0.9 &  3.6 & g&  7.65 &   1.59 &   8.71 &   9.04 \\
028593 & 5.28 & 0.73 & 5516 & 4.55 & 0.9 &  4.5 & g&  7.67 &   1.36 &   8.69 &   9.06 \\
031609 & 5.36 & 0.74 & 5508 & 4.55 & 0.9 &  3.7 & g&  7.67 &   1.61 &   8.73 &   8.97 \\
027250 & 5.48 & 0.75 & 5485 & 4.55 & 0.9 &  4.2 & g&  7.59 &   1.49 &   8.75 & (8.90)\\
027732 & 5.41 & 0.76 & 5449 & 4.55 & 0.9 &  4.3 & g&  7.61 &   1.31 &   8.76 & (8.86)\\
032347 & 5.33 & 0.76 & 5429 & 4.56 & 0.8 &  4.8 & g&  7.71 &   1.49 &   8.84 &   9.21 \\
242780 & 5.34 & 0.76 & 5429 & 4.56 & 0.8 &  4.8 & g&  7.69 &   1.40 &   8.77 & (9.05)\\
283704 & 5.35 & 0.77 & 5426 & 4.56 & 0.8 &  3.7 & g&  7.72 &   1.36 & (8.85)& (8.70)\\
284574 & 5.42 & 0.81 & 5303 & 4.57 & 0.8 &  4.8 & g&  7.76 &   1.04 & (8.93)& (8.85)\\
284253 & 5.59 & 0.81 & 5297 & 4.57 & 0.8 &  3.6 & g&  7.66 &  (0.75)& (8.84)& (9.03)\\
285773 & 5.87 & 0.83 & 5254 & 4.58 & 0.7 &  3.8 & g&  7.73 &  (0.75)& (8.67)& (9.32)\\
030505 & 5.63 & 0.83 & 5249 & 4.58 & 0.7 &  4.4 & g&  7.74 &  (0.57)& (8.83)& (9.22)\\
028258 & 5.66 & 0.84 & 5235 & 4.58 & 0.7 &  4.1 & g&  7.72 &  (0.29)& (8.87)&   9.21 \\
027771 & 5.74 & 0.86 & 5196 & 4.58 & 0.7 &  4.5 & g&  7.78 &  (0.50)& (8.91)&   9.07 \\
028462 & 6.04 & 0.87 & 5172 & 4.58 & 0.7 &  4.5 & g&  7.75 &  (0.82)& (8.98)& (8.55)\\
285367 & 5.79 & 0.89 & 5114 & 4.59 & 0.7 &  4.3 & g&  7.70 &  (0.10)& (8.94)& (9.06)\\
285252 & 5.91 & 0.89 & 5103 & 4.59 & 0.6 &  4.2 & g&  7.79 & ($-0.29$)& (9.03)& (8.32)\\
\end{longtable}

\clearpage
\setcounter{table}{1}
\footnotesize
\begin{table}[h]
\caption{Atomic parameters of important lines relevant for spectrum fitting.}
\begin{center}
\begin{tabular}{ccrcc}\hline\hline
Species & $\lambda$ & $\chi$ & $\log gf$ & Remark\\
\hline
\multicolumn{4}{c}{[6080--6089~$\rm\AA$ fitting]}\\
V~{\sc i} &6081.441 & 1.05 & $-$0.58 &  \\
Co~{\sc i}&6082.422 & 3.51 & $-$0.52 &  \\
Fe~{\sc i}&6082.708 & 2.22 & $-$3.57 &  \\
Fe~{\sc ii}&6084.111& 3.20 & $-$3.81 &  \\
Ti~{\sc i}&6085.228 & 1.05 & $-$1.35 &  \\
Fe~{\sc i}&6085.260 & 2.76 & $-$3.21 &  \\
Ni~{\sc i}&6086.276 & 4.27 & $-$0.53 &  \\
Co~{\sc i}&6086.658 & 3.41 & $-$1.04 &  \\
Si~{\sc i}&6087.805 & 5.87 & $-$1.60 &  \\
\hline
\multicolumn{4}{c}{[6703--6709~$\rm\AA$ fitting]}\\
Fe~{\sc i} & 6703.568 & 2.76 & $-$3.02 & (adjusted)\\
Fe~{\sc i} & 6705.101 & 4.61 & $-$1.02 & (adjusted)\\
Fe~{\sc i} & 6707.441 & 4.61 & $-$2.35 & \\
Li~{\sc i} & 6707.756 & 0.00 & $-$0.43 & Li~6708\\ 
Li~{\sc i} & 6707.768 & 0.00 & $-$0.21 & Li~6708\\
Li~{\sc i} & 6707.907 & 0.00 & $-$0.93 & Li~6708\\
Li~{\sc i} & 6707.908 & 0.00 & $-$1.16 & Li~6708\\
Li~{\sc i} & 6707.919 & 0.00 & $-$0.71 & Li~6708\\
Li~{\sc i} & 6707.920 & 0.00 & $-$0.93 & Li~6708\\
\hline
\multicolumn{4}{c}{[7110--7121~$\rm\AA$ fitting]}\\
Ni~{\sc i} &7110.892 & 1.94 & $-$2.88 & (adjusted)\\
 C~{\sc i} &7111.472 & 8.64 & $-$1.24 & (adjusted)\\
Fe~{\sc i} &7112.168 & 2.99 & $-$2.89 & (adjusted)\\
 C~{\sc i} &7113.178 & 8.65 & $-$0.80 & (adjusted),C~7113\\
 C~{\sc i} &7115.172 & 8.64 & $-$0.96 & (adjusted)\\
 C~{\sc i} &7116.991 & 8.65 & $-$0.91 & \\
Fe~{\sc i} &7118.119 & 5.01 & $-$1.39 & (adjusted)\\
 C~{\sc i} &7119.656 & 8.64 & $-$1.13 & (adjusted)\\
Fe~{\sc i} &7120.022 & 4.56 & $-$1.91 & (adjusted)\\
\hline
\multicolumn{4}{c}{[6156--6159~$\rm\AA$ fitting]}\\
O~{\sc i} &6156.737 &10.74 & $-$1.52 &  \\
O~{\sc i} &6156.755 &10.74 & $-$0.93 &  \\
O~{\sc i} &6156.778 &10.74 & $-$0.73 &  \\
Fe~{\sc i}&6157.725 & 4.08 & $-$1.26 &  \\
O~{\sc i} &6158.149 &10.74 & $-$1.89 & O~6158  \\
O~{\sc i} &6158.172 &10.74 & $-$1.03 & O~6158  \\
O~{\sc i} &6158.187 &10.73 & $-$0.44 & O~6158  \\
\hline
\end{tabular}
\end{center}
\footnotesize
Note. 
$\lambda$ is the air wavelength (in $\rm\AA$), $\chi$ is the lower excitation 
potential (in eV), and $\log gf$ is the logarithm of $g$ (statistical weight
of the lower level) times $f$ (absorption oscillator strength).
These data were taken primarily from the compilation of Kurucz and Bell (1995),
though empirically adjusted ``solar $gf$ values'' were applied in several cases
(remarked as ``adjusted'' in column 5). Regarding lithium, we considered only
the component lines of $^{7}$Li, neglecting those of $^{6}$Li.
\end{table}

\end{document}